\documentclass[journal=mamobx,manuscript=article]{achemso}
\setkeys{acs}{articletitle=true}

\SectionNumbersOn

\usepackage[version=3]{mhchem} 
\usepackage[T1]{fontenc} 

\usepackage{color}
\usepackage{gensymb}
\usepackage {indentfirst}

\author{Yue-Tong Dong}
\affiliation{State Key Laboratory of Polymer Science and Technology, Changchun Institute of Applied Chemistry, Chinese Academy of Sciences, Changchun 130022, P. R. China}
\alsoaffiliation{School of Applied Chemistry and Engineering, University of Science and Technology of China, Hefei 230026, P. R. China}

\author{Jack F. Douglas}
\altaffiliation{NIST Fellow (emeritus).}
\affiliation{Materials Science and Engineering Division, National Institute of Standards and Technology, Gaithersburg, Maryland 20899, United States}

\author{Wen-Sheng Xu}
\email{wsxu@ciac.ac.cn}
\affiliation{State Key Laboratory of Polymer Science and Technology, Changchun Institute of Applied Chemistry, Chinese Academy of Sciences, Changchun 130022, P. R. China}
\alsoaffiliation{School of Applied Chemistry and Engineering, University of Science and Technology of China, Hefei 230026, P. R. China}

\title{Tunable Cooperative Motion, Rigidity, and Glassy Dynamics in Knotted Ring Polymer Melts}

\abbreviations{IR,NMR,UV}

\begin{document}
	
\begin{tocentry}
		
 \centering
 \includegraphics[height=2.5cm]{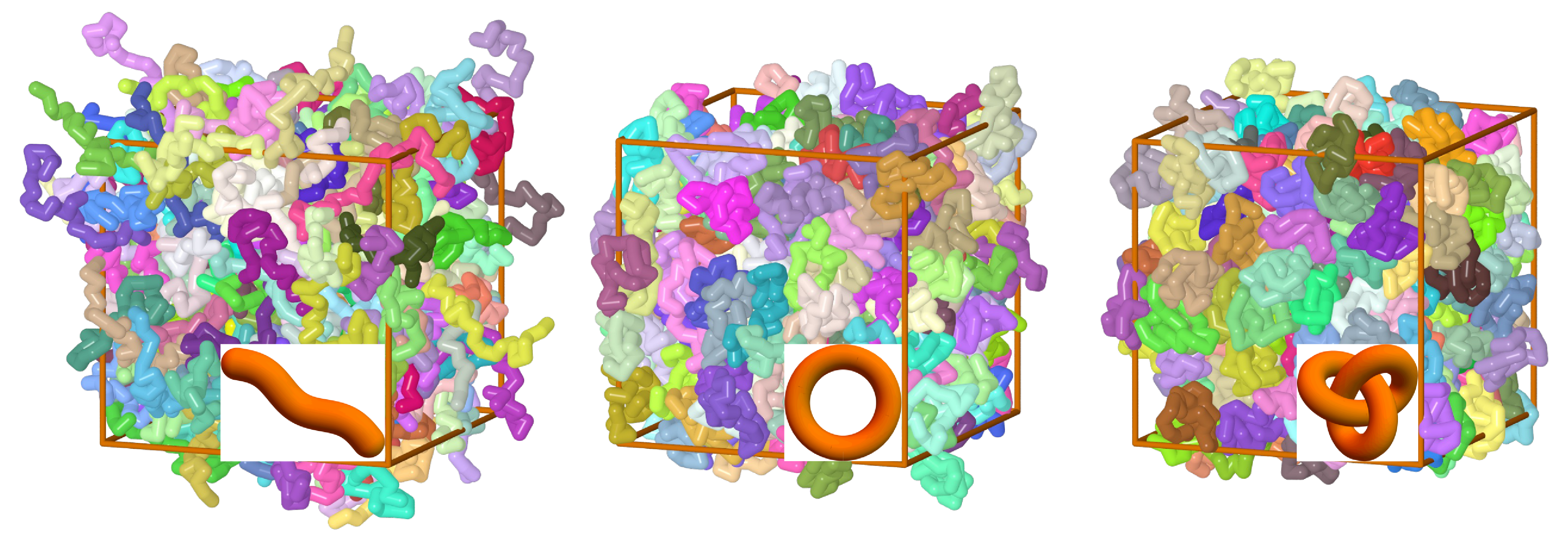}

	
\end{tocentry}
	
\newpage
	
\begin{abstract}

We present a molecular dynamics study of the influence of knot complexity and molecular mass on glass formation upon cooling in knotted ring polymer melts. We find that cooperative motion, rigidity, and glassy dynamics can be tuned over a wide range by knots. By leveraging these knotting constraints, we assess the validity of prevalent models of glass formation, including the string model based on cooperative particle motion, the localization model emphasizing fluctuations in local particle mobility, and the shoving model derived from emergent elastic properties in relation to material stiffness. In line with our previous findings on polymeric and other glass-forming liquids, we demonstrate that all these models of glass formation provide a quantitative description of segmental relaxation as a function of knot complexity, molecular mass, and temperature, despite their apparently distinct conceptual foundations. Our study thus provides additional evidence for an underlying unity among various theoretical frameworks and for the presence of quantitative relations between the characteristic properties emphasized by these models. Furthermore, we discuss dynamic and elastic heterogeneities in relation to fragility and stiffness variations of knotted ring polymer melts, with a focus on how these trends relate to other glass-forming liquids where fragility is tuned over a large range.

\end{abstract}

\newpage

\section{\label{Sec_Intro}Introduction}

Glass-forming materials exhibit a dramatic slowdown in their dynamics upon cooling toward the glass transition temperature ($T_\mathrm{g}$), despite the absence of significant changes in their overall static structure. \cite{50th_2017_50_6333, Supercooled_2001_410_259, Theoretical_2011_83_587, Perspective_2012_137_080901, Perspective_2013_138_12A301, Glass_2013_4_263} This complex phenomenon has motivated the development of a number of theoretical approaches based on distinct physical perspectives, such as local density-based free volume models, \cite{Molecular_1959_31_1164, Free_1961_34_120} the Adam-Gibbs (AG) theory, \cite{Temperature_1965_43_139} the mode-coupling theory, \cite{Book_Gotze, Dynamical_1984_29_2765, Dynamics_1984_17_5915} the random first-order transition theory, \cite{Spin_1988} the string model, \cite{Communication_2014_141_141102, String_2014_140_204509} the localization model, \cite{Generalized_2012_8_11455, Localization_2016_2016_054048} the shoving model, \cite{Colloquium_2006_78_953, instantaneous_2012_136_224108, review_2015_407_14} and a continually growing number of other approaches. These diverse models have yielded valuable insights into specific aspects of glass formation and stimulated extensive experimental and computational efforts to test their assumptions and predictions, \cite{50th_2017_50_6333, Supercooled_2001_410_259, Theoretical_2011_83_587, Perspective_2012_137_080901, Perspective_2013_138_12A301, Glass_2013_4_263} although a universally accepted theory of glass formation has not been established.

While the growing number of proposed models might be viewed as representing a disappointing lack of consensus on the nature of glass formation, their independent success in describing the dynamics of glass-forming materials suggests to us a possible unity underlying these models. Indeed, there has been progress made in this direction recently. Betancourt et al. \cite{Quantitative_2015_112_2966} presented evidence from simulations of polymeric glass-forming liquids in the bulk, thin supported films, and polymer nanocomposites that relaxation can be described reasonably well for all of the systems by different models based on cooperative motion, emergent elasticity, and dynamical free volume. Following this line of research, our recent works demonstrated that some of the existing models provide equally valid descriptions of glass formation in polymer melts having variable chain length and chain rigidity \cite{Parallel_2023_56_4929} as well as in the canonical Kob-Andersen model system over a wide range of densities and pressures. \cite{Understanding_2024_128_10999} Moreover, the generalized entropy theory (GET) of polymer glass formation, \cite{Generalized_2008_137_125, Polymer_2021_54_3001, Thermodynamic_2023_41_1329, Advances_2023_53_616} which combines the lattice cluster theory \cite{Lattice_1998_103_335, Lattice_2014_141_044909} for polymer thermodynamics and the AG relation \cite{Temperature_1965_43_139} relating the configurational entropy to the structural relaxation time, predicts close interrelations between the configurational entropy, configurational enthalpy, and configurational internal energy, \cite{Thermodynamic_2022_55_8699} bringing together these disparate thermodynamic viewpoints of glass formation. \cite{Quantitative_2018_2_055604, Quantitative_2019_52_1424, Modelfree_2020_152_044901} The reader is referred to a recent review for the detailed discussion of a unified thermodynamic perspective on glass formation. \cite{Thermodynamic_2023_41_1329}

In the present work, we extend our previous efforts in unifying various models of glass formation to a family of knotted ring polymer melts. Knots have been found in various polymeric systems, such as Deoxyribonucleic acid (DNA), \cite{Probability_1993_90_5307, Knotting_1993_260_533, Fractal_2007_98_058102} proteins, \cite{Deeply_2000_406_916, KnotProt_2015_43_D306, Tightening_2009_96_1508} and synthetic polymers. \cite{Knots_2005_127_15102, Knots_2021_54_5321, Knotting_2021_33_244001} It has been widely appreciated that the physical properties of polymeric materials can significantly be impacted by knots, such as the tensile strength, \cite{Influence_1999_399_46} the relaxation rate of compressed polymers, \cite{Compression_2011_108_16153} the diffusion of polymers in nanotubes, \cite{Translocation_2008_129_121107, Topological_2012_109_118301} the dynamics and functionality of DNA and proteins, \cite{Stretching_2015_11_3105, Dynamics_2017_50_4074, Biophysics_2010_39_349} etc. Knotted ring polymers also serve as an ideal system for testing theoretical models of glass formation, since knotting can lead to large alterations in the characteristic properties of glass formation, such as collective motion and material stiffness. In our recent work based on extensive simulation results, \cite{Glass_2024_57_6875} we examined the influence of knot complexity (quantified by the minimum crossing number $m_c$) and molecular mass on basic thermodynamic and segmental dynamic properties in knotted ring polymer melts, and inspired by the GET predictions, \cite{Generalized_2008_137_125, Polymer_2021_54_3001} we rationalized the trends in $T_\mathrm{g}$ and fragility in terms of molecular packing frustration, where fragility is a property that characterizes the strength of the $T$ dependence of the dynamics near $T_{\mathrm{g}}$. Hence, leveraging these knotting constraints offers a promising avenue to illuminate longstanding theoretical questions concerning glass formation. As a particular advantage of studying glass-forming knotted ring polymers, and also star polymers, \cite{Confinement_2024_160_044503} the characteristics of glass formation such as $T_\mathrm{g}$ and fragility can be varied over a large range without changing the monomer structure or molecular mass. Moreover, one can avoid phase separation and crystallization tendencies in popular glass-forming multi-component fluid mixtures, such as the Kob-Andersen model system. \cite{Testing_1995_51_4626, Testing_1995_52_4134} From a broader perspective, a deep understanding of how knotting influences glass formation provides important guidance for designing advanced polymeric materials with desired properties \cite{Knotting_2022_51_7779, What_2024_6_2084, Hierarchical_2025_135_248201} and holds direct relevance to understanding the physical properties and biological functions of topologically complex biomacromolecules. \cite{Allosteric_2016_352_1555, Methyl_2016_23_941, Folding_2023_83_102709}

After a brief overview of relaxation in knotted ring polymer melts having variable knot complexity and molecular mass, we assess the validity of several models of glass formation. Specifically, we consider the string model, \cite{Communication_2014_141_141102, String_2014_140_204509} a variant of the entropy theory of glass formation emphasizing cooperative particle motion, the localization model emphasizing fluctuations in local particle mobility, \cite{Generalized_2012_8_11455, Localization_2016_2016_054048} and the shoving model derived from a focus on the emergent elastics of glass-forming liquids. \cite{Colloquium_2006_78_953, instantaneous_2012_136_224108, review_2015_407_14} We demonstrate that all these models of glass formation provide a quantitative description of segmental relaxation as a function of knot complexity, molecular mass, and temperature, despite their apparently distinct conceptual foundations. We then show that the extent of collective motion and the high frequency shear modulus in knotted ring polymer melts are closely interrelated, following the empirical relations established in our previous works based on linear polymer melts \cite{Parallel_2023_56_4929} and the Kob-Andersen model system. \cite{Understanding_2024_128_10999} Finally, we discuss dynamic and elastic heterogeneities in relation to fragility and stiffness variations of knotted ring polymer melts, with a focus on how these trends relate to other glass-forming liquids where fragility is tuned over a large range.

\section{\label{Sec_Method}Method}

\begin{figure*}[htb!]
	\centering
	\includegraphics[angle=0, width=0.8\textwidth]{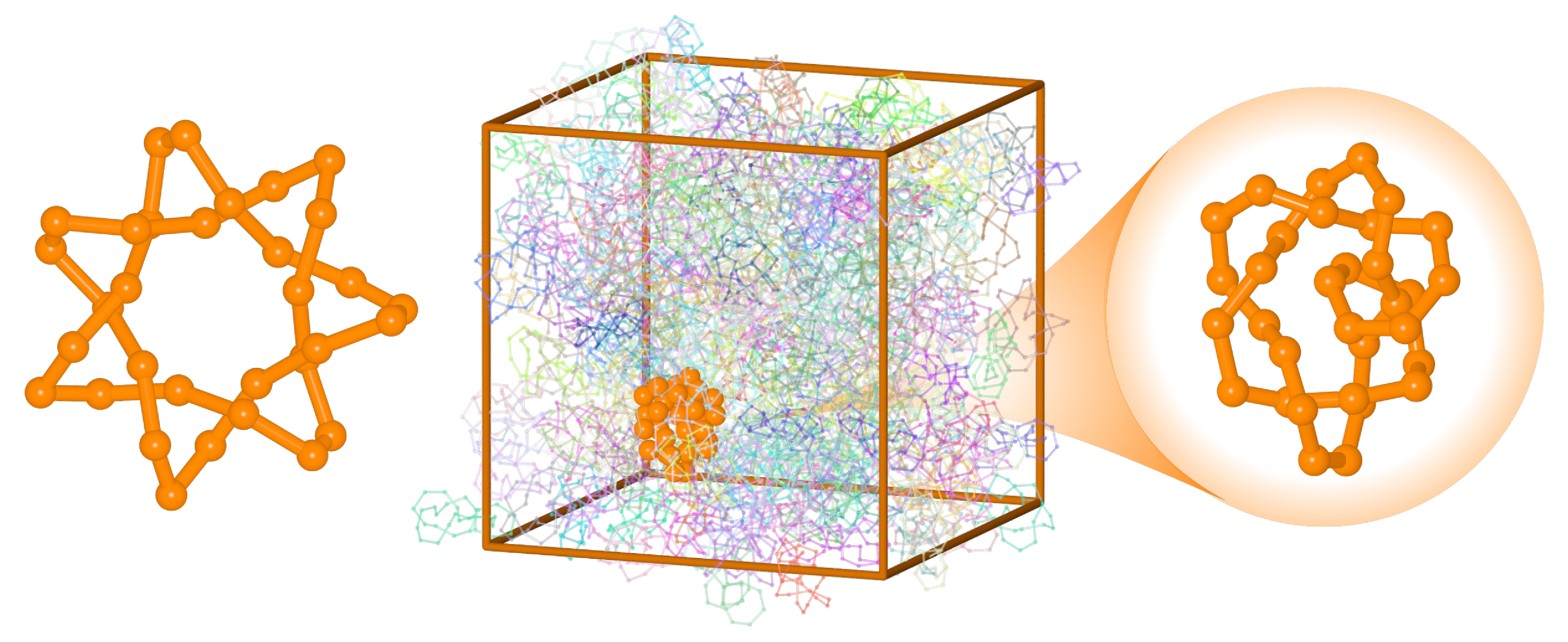}
	\caption{Example of a knotted ring polymer melt with the minimum crossing number of $m_c = 7$ and the molecular mass of $M = 32$. The left part illustrates the initial configuration of a knotted ring polymer. The right part displays a melt composed of $250$ chains after equilibration, where one of the chains is highlighted and all the others are shown as translucent thin cylinders.}
	\label{Fig_Schematic}
\end{figure*}

Our computational study is based on a generic, coarse-grained model of polymer melts. \cite{Dynamics_1990_92_5057, Molecular_1986_33_3628} Since the model and simulation details have been provided at length in our previous work, \cite{Glass_2024_57_6875} our description is brief here. Our simulated polymer melt is composed of $N_p$ knotted rings of molecular mass $M$ and knot complexity $m_c$, where $M$ is the number of beads in a single chain and $m_c$ is the minimum crossing number. $N_p$ is adjusted for different $M$ so that the total number of beads, $N = N_p M$, ranges from $8000$ to $12800$. The method for generating initial knot configurations is consistent with that of Vargas-Lara et al., \cite{Influence_2019_150_101103} which employs Mathematica \cite{Mathematica} to specify representative configurations for a prescribed knot topology (Figure \ref{Fig_Schematic}).

The present study deliberately selected prime knots with the subscript $1$ (e.g., $3_1$, $4_1$, $5_1$, $6_1$, and $7_1$) as research objects because they possess the highest symmetry within their respective crossing number classes. This high degree of symmetry makes them ideal benchmarks for establishing fundamental physical models. Notably, our selection already encompasses both torus (e.g., $3_1$, $5_1$, and $7_1$) and twist knots (e.g., $4_1$ and $6_1$), which are the most representative knot families in polymer physics. While more subtle topological features, such as knot chirality or non-prime configurations, may introduce measurable differences in glass formation, our current findings suggest that the minimum crossing number $m_c$ serves as the dominant descriptor for the observed ``topological rigidification'' across these different families. The exploration of how more complex topological details impact glass formation remains an interesting direction for future work.

The force field comprises two potentials. Nonbonded beads interact via a truncated-and-shifted Lennard-Jones (LJ) potential with an energy parameter $\varepsilon$, a length parameter $\sigma$, and a cutoff distance of $r_c = 2.5 \sigma$. Chain connectivity is maintained by the finitely extensible nonlinear elastic (FENE) spring potential \cite{Dynamics_1990_92_5057, Molecular_1986_33_3628} between adjacent beads, with a spring constant of $k_b = 30 \varepsilon / \sigma^{2}$ and a maximum extension of $R_0 = 1.5 \sigma$. All quantities are reported in standard reduced LJ units; i.e., length, energy, temperature, and pressure are expressed in units of $\sigma$, $\varepsilon$, $\varepsilon / k_\mathrm{B}$, and $\varepsilon / \sigma^{3}$, respectively, where $k_\mathrm{B}$ is the Boltzmann constant. In addition, we define the time unit by $\tau = \sqrt{m_b\sigma^{2} / \varepsilon}$, where $m_b$ is the bead mass. All our simulations were performed using the Large-scale Atomic/Molecular Massively Parallel Simulator (LAMMPS) package \cite{LAMMPS_2022_271_108171, Lammps} at zero pressure using a time step of $\Delta t = 0.005 \tau$. Temperature and pressure were controlled via the Nos\'e-Hoover thermostat and barostat. Prior to data collection, the system was equilibrated for a duration significantly longer than its segmental structural relaxation time $\tau_{\alpha}$. Figure \ref{Fig_Schematic} shows an example of a knotted ring polymer melt composed of $250$ chains with $m_c = 7$ and $M = 32$ after equilibration. We confirmed via an Alexander polynomial-based approach of Dai and coworkers \cite{Simple_2024_42_2030} that the knot types generated initially remain unchanged after equilibration in our simulations. We refer the reader to ref \citenum{Glass_2024_57_6875} for additional technical details.

\section{Results and Discussion}

\subsection{\label{Sec_Overview}Overview of Structural Relaxation Time}

Before assessing the validity of various models of glass formation in knotted ring polymer melts, we first review how basic relaxation properties are influenced by knot complexity and molecular mass. The quantity of primary interest here is the segmental structural relaxation time $\tau_{\alpha}$, which is obtained from the time ($t$) dependence of the self-intermediate scattering function $F_s(q, t)$. The wave number of $q^* = q = 7.0 \sigma^{-1}$ is chosen to correspond to the first peak of the static structure factor $S(q)$ and thus to the typical intersegmental distance. $\tau_{\alpha}$ is then defined by the time at which $F_s(q, t)$ decays to $0.1$. \cite{Glass_2024_57_6875} $F_s(q, t)$ also enables the definition of the non-ergodicity parameter $f_{s, q^*}$ from $f_{s, q^*} = F_s(q, t = 1 \tau)$, \cite{Parallel_2023_56_4929} another quantity of interest, by adopting the caging onset time of $t_{\mathrm{onset}} = 1 \tau$. The specific definition of $F_s(q,t)$ is given in Section S1 of the Supporting Information, along with typical examples.
 
\begin{figure*}[htb!]
	\centering
	\includegraphics[angle=0, width=0.8\textwidth]{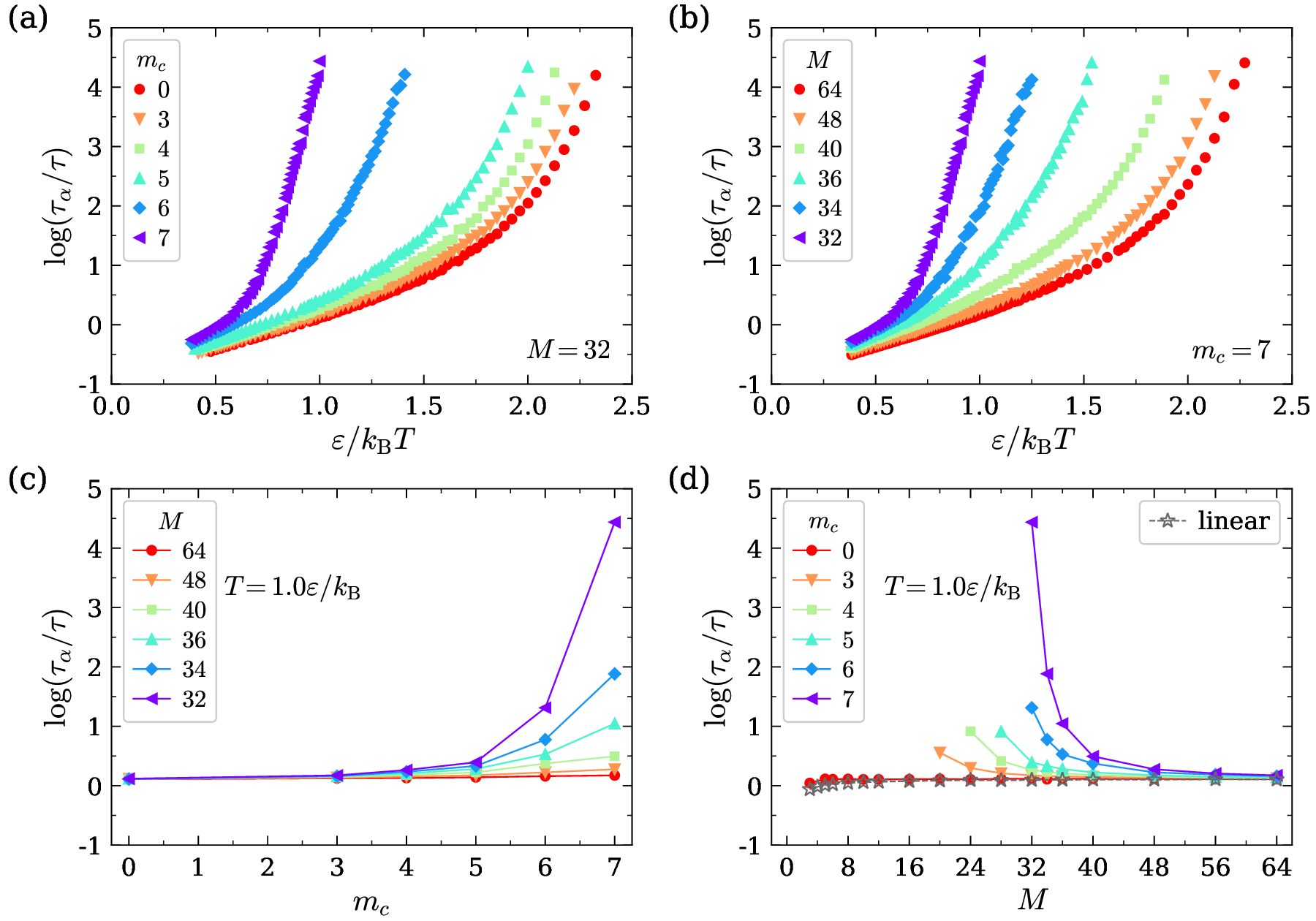}
	\caption{Structural relaxation time $\tau_{\alpha}$ in variation with temperature $T$, knot complexity $m_c$, and molecular mass $M$. Panels (a) and (b) show $\log (\tau_{\alpha} / \tau)$ versus $\varepsilon / k_{\mathrm{B}} T$ for a range of $m_c$ at $M = 32$ and for a range of $M$ at $m_c = 7$, respectively. Panels (c) and (d) show the variations with $m_c$ and $M$ of $\log (\tau_{\alpha} / \tau)$ over a range of fixed $M$ and $m_c$ at $T = 1.0 \varepsilon / k_{\mathrm{B}}$, respectively. The results for linear polymer melts are included as a reference in panel (d).}
	\label{Fig_Tau}
\end{figure*}

To frame our subsequent investigation below, Figure \ref{Fig_Tau} summarizes the representative results for the variations of $\tau_{\alpha}$ with $T$, $m_c$, and $M$. The $T$ dependence of $\log (\tau_{\alpha})$ is shown for a range of $m_c$ at $M = 32$ and for a range of $M$ at $m_c = 7$ in Figure \ref{Fig_Tau}a, b, respectively. We see that segmental dynamics slows down dramatically upon cooling and that both the magnitude and $T$ dependence of $\tau_{\alpha}$ can be affected strongly by knot complexity and molecular mass. Accordingly, our previous paper provided a detailed discussion of the variation of $\tau_{\alpha}$ with $m_c$ and $M$, \cite{Glass_2024_57_6875} along with the characteristic temperatures (such as $T_{\mathrm{g}}$) and fragility of glass formation, as determined from the $T$ dependence of $\tau_{\alpha}$. To examine the influence of knot complexity and molecular mass on $\tau_{\alpha}$ at fixed $T$, Figure \ref{Fig_Tau}c indicates that $\tau_{\alpha}$ initially changes with $m_c$ weakly and then increases abruptly with further increasing $m_c$ at about $m_c = 5$ in the low $M$ regime, in line with the simulation results of Vargas-Lara et al. \cite{Influence_2019_150_101103} At sufficiently large $M$, however, knotting effects on $\tau_{\alpha}$ are quite weak over the entire range of $m_c$ considered in the present work. Concerning the $M$ dependence of $\tau_{\alpha}$, a previous work based on the same coarse-grained model showed that $\tau_{\alpha}$ increases with increasing $M$ in linear polymer melts at low $T$, while $\tau_{\alpha}$ changes rather weakly with $M$ in unknotted ring polymer melts. \cite{Understanding_2023_41_1447} In the presence of knots, however, $\tau_{\alpha}$ is found to increase with decreasing $M$ (Figure \ref{Fig_Tau}d). Indeed, our simulation results suggest that $\tau_{\alpha}$ tends to diverge at a critically small $M$, a phenomenon that we plan to investigate thoroughly elsewhere. Notice that the temperature $T = 1.0 \varepsilon / k_{\mathrm{B}}$ is essentially in the Arrhenius regime at high $T$ for linear polymer melts so that $M$ dependence of $\tau_{\alpha}$ is observed to be rather weak in Figure \ref{Fig_Tau}d.

The singular increase of $\tau_\alpha$ with decreasing $M$ at fixed $m_c$ (Figure \ref{Fig_Tau}d) is particularly noteworthy. Our simulation results suggest that $\tau_\alpha$ tends to diverge at a critically small $M$, a phenomenon that implies a transition toward a topologically induced dynamical arrest. We find that the $M$ dependence of $\tau_{\alpha}$ at fixed $T$ can be reasonably described by an analogous Vogel-Fulcher-Tammann (VFT) functional form, $\tau_\alpha = \tau_0 \exp[D_M M_0 / (M - M_0)]$, where $\tau_0$ is a prefactor, $D_M$ quantifies the strength of the $M$ dependence of $\tau_\alpha$, and $M_0$ is the molecular mass where $\tau_\alpha$ extrapolates to infinity. This result implies that as the polymer chain length approaches the geometric limit required to maintain the knot, a topology-induced glass formation occurs. Evidently, this significant phenomenon is worthy of an extended discussion. This discussion requires us to carry out extended simulations at smaller $M$ than those considered in the present work, where we focus on the knotted ring polymer melts upon cooling, a common route to glass formation. We plan to provide a more systematic scaling analysis and a thorough investigation of this critically small $M$ regime in future work.

In previous works, \cite{Influence_2019_150_101103, Glass_2024_57_6875} the large variation of $\tau_{\alpha}$ with $m_c$ was attributed to ``topological rigidification'' induced by knots in the sense that chains are stiffened by knotting. Our previous systematic investigation of the conformational properties of knotted ring polymer melts \cite{Glass_2024_57_6875} demonstrated that the persistence length and other size-related metrics undergo significant changes with varying $m_c$ and $M$, providing a direct structural foundation for this mechanism. While such an idea is useful for understanding the general trends qualitatively, it remains unknown if a quantitative description of the segmental relaxation in knotted ring polymer melts can be obtained based on the prevailing models of glass formation initially proposed for simpler systems such as the Kob-Andersen model system and linear polymer melts. Below, we discuss the effect of knotting on the extent of collective motion and material stiffness and examine the role of these properties in the quantitative description of the segmental structural relaxation time. This investigation is particularly worthwhile, given the fact that both $T_{\mathrm{g}}$ and fragility can be tuned over a large range by varying knot complexity and molecular mass. \cite{Glass_2024_57_6875} In particular, we found that the characteristic temperatures of glass formation of knotted ring polymer melts, including $T_{\mathrm{g}}$, increase progressively with increasing $m_c$, while the fragility of glass formation decreases progressively with increasing $m_c$ under the condition of relatively small $M$. The decrease of fragility with increasing $m_c$ was accompanied by a reduction of the dimensionless thermal expansion coefficient and isothermal compressibility, a general trend predicted by the GET. \cite{Polymer_2021_54_3001}

\subsection{\label{Sec_String}String Model Description of Glass Formation}

Dynamic heterogeneity is a universal feature of glass-forming liquids upon cooling toward $T_\mathrm{g}$. This phenomenon is characterized by the formation of spatially correlated regions with markedly different dynamics in that some regions are composed of relatively mobile or fast particles, while particles in other complementary regions display significantly lower mobility. This widely observed phenomenon is also regarded as being the key to understanding the anomalous physical properties of glass-forming materials. Moreover, extensive simulations \cite{Stringlike_1998_80_2338, Polymerspecific_2003_119_5290, Modifying_2011_106_115702, Fragility_2013_9_241, Relationship_2013_138_12A541, Interfacial_2014_5_4163, String_2014_140_204509, Quantitative_2015_112_2966, Unifying_2015_142_234907} have revealed that the most mobile particles mover cooperatively in a stringlike fashion. Such stringlike cooperative motion has been shown to provide a quantitative realization of the hypothetical ``cooperatively arranging regions'' (CRR) in the AG model. \cite{Temperature_1965_43_139}

\begin{figure*}[htb!]
	\centering
	\includegraphics[angle=0, width=0.8\textwidth]{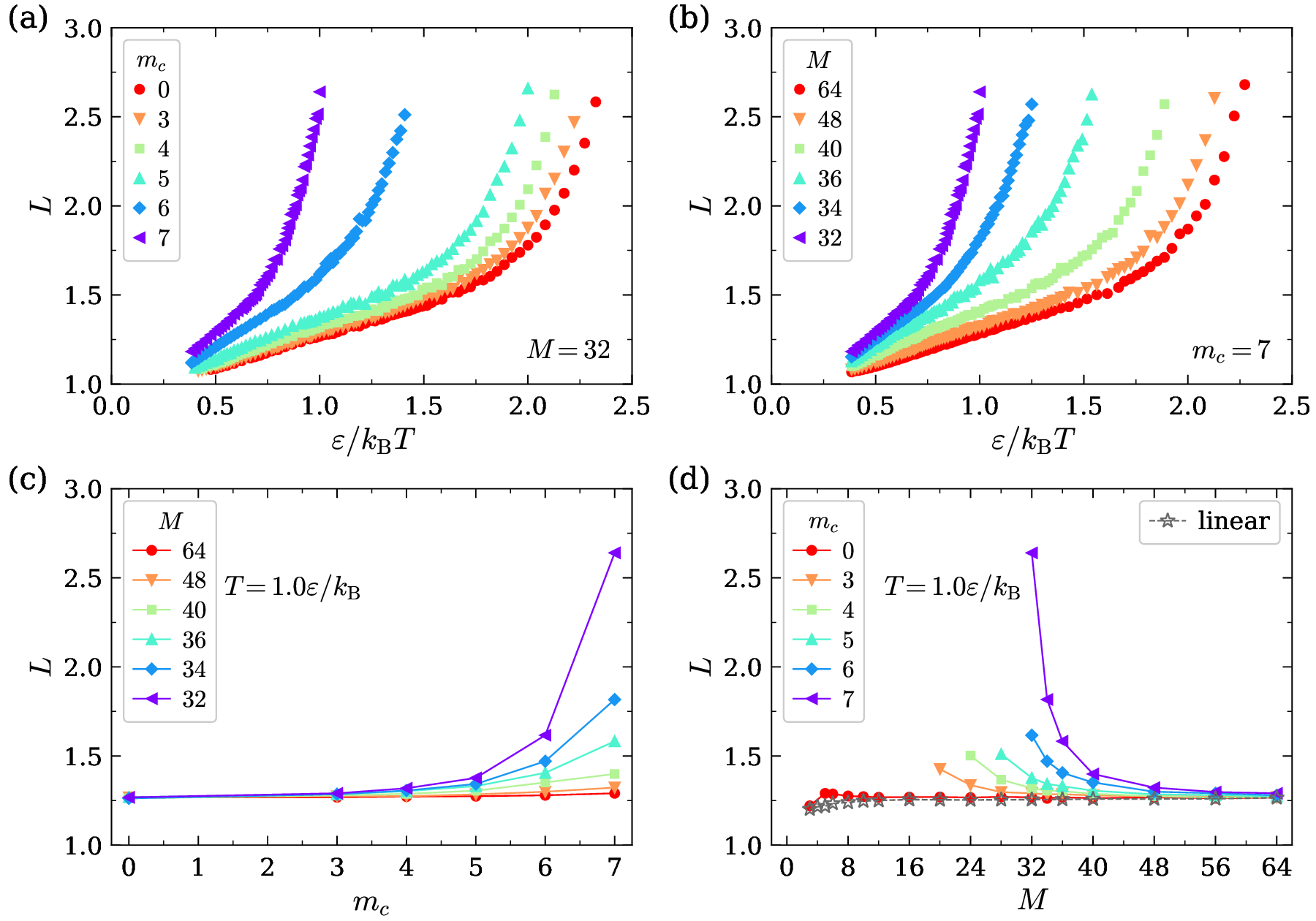}
	\caption{Average string length $L$ in variation with temperature $T$, knot complexity $m_c$, and molecular mass $M$. Panels (a) and (b) show $L$ versus $\varepsilon / k_{\mathrm{B}} T$ for a range of $m_c$ at $M = 32$ and for a range of $M$ at $m_c = 7$, respectively. Panels (c) and (d) show the variations with $m_c$ and $M$ of $L$ over a range of fixed $M$ and $m_c$ at $T = 1.0 \varepsilon / k_{\mathrm{B}}$, respectively. The results for linear polymer melts are included as a reference in panel (d).}
	\label{Fig_ASL}
\end{figure*}

We follow the established procedure for polymeric systems to quantify the stringlike cooperative motion and determine the average string length $L$. \cite{Polymerspecific_2003_119_5290, Relationship_2013_138_12A541} Details can be found in Section S1 of the Supporting Information. Mirroring our discussion of $\tau_{\alpha}$ in Section \ref{Sec_Overview}, Figure \ref{Fig_ASL} summarizes the representative results for the variations of $L$ with $T$, $m_c$, and $M$. We find that $L$ exhibits similar trends to those of $\tau_{\alpha}$. The $T$ dependence of $L$ is presented for a range of $m_c$ at $M = 32$ and for a range of $M$ at $m_c = 7$ in Figure \ref{Fig_ASL}a, b. These results demonstrate that both $m_c$ and $M$ can likewise influence $L$ quite strongly. Specifically, Figure \ref{Fig_ASL}c indicates that $L$ varies weakly with $m_c$ when $m_c$ is small, but significant changes occur when $m_c$ is increased to $5$ in the low $M$ regime. However, when $M$ is large, the variation of $L$ with $m_c$ remains weak across the entire range of $m_c$ considered. Additionally, Figure \ref{Fig_ASL}d indicates that $L$ gradually increases as $M$ decreases, which is different from the trends for linear and unknotted ring polymer melts. These results imply that the extent of cooperative motion can be tuned by knots over a wide range.

The string model of glass formation is built upon the finding that stringlike cooperative motion $L$ provides a reasonable realization of the abstract CRR in the AG model. \cite{Temperature_1965_43_139} The validity of the string model has been supported by studies based on a range of polymeric and other glass-forming systems. \cite{Interfacial_2014_5_4163, String_2014_140_204509, Quantitative_2015_112_2966, Unifying_2015_142_234907, Influence_2016_49_8355, Stringlike_2016_5_1375, Influence_2017_50_2585, Collective_2019_123_5935, Dynamic_2020_152_054904, Role_2015_142_164506} The string model finds further theoretical justification in the analysis of Freed, \cite{Communication_2014_141_141102} who extended transition state theory (TST) \cite{theory_1941_28_301, Book_Eyring} to account for such stringlike cooperative barrier-crossing events in glass-forming liquids.

\begin{figure*}[htb!]
	\centering
	\includegraphics[angle=0, width=0.8\textwidth]{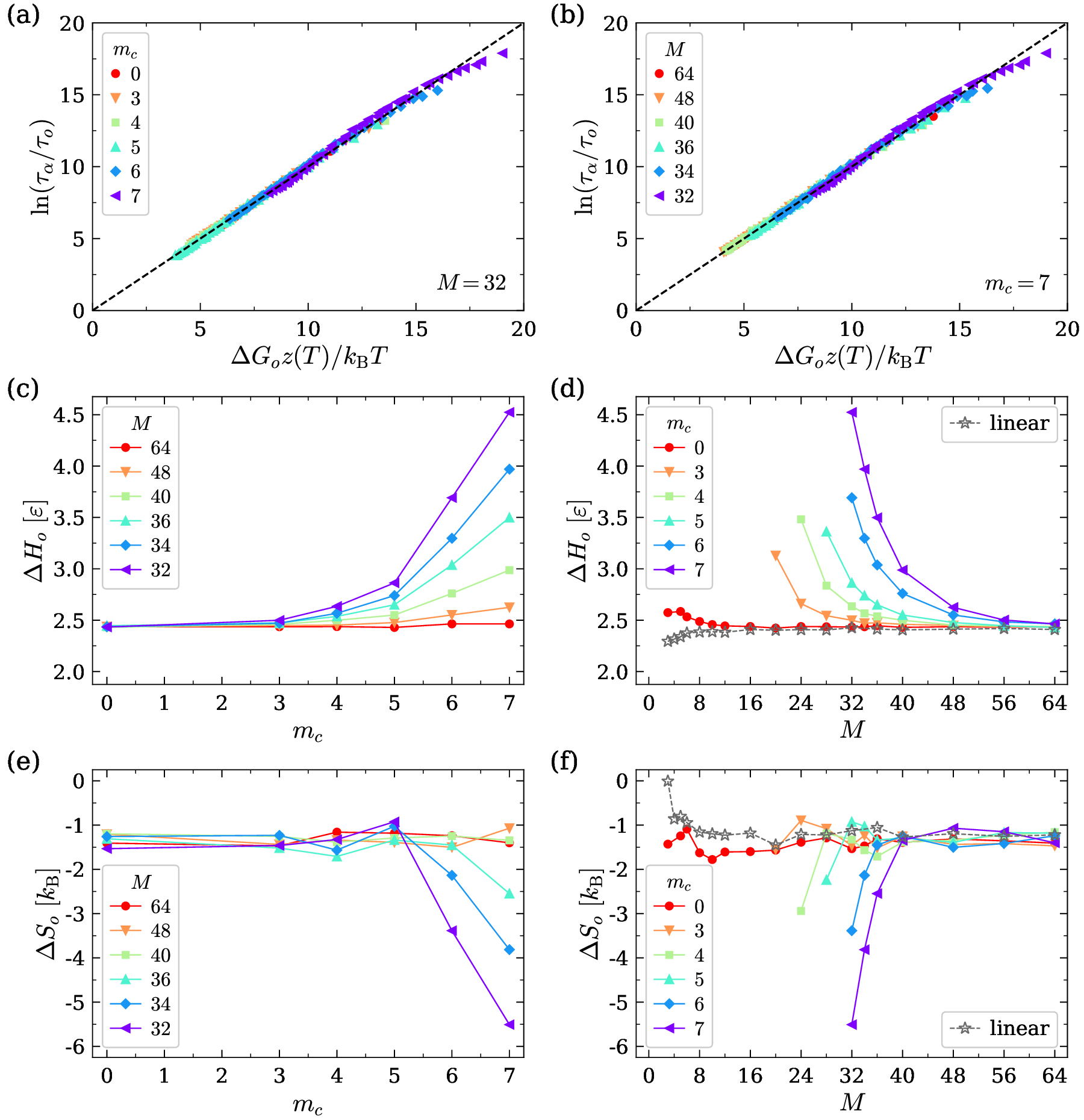}
	\caption{String model description of glass formation. Panels (a) and (b) show $\ln (\tau_{\alpha} / \tau_o)$ versus $\Delta G_o z(T) / k_{\mathrm{B}} T$ for a range of $m_c$ at $M = 32$ and for a range of $M$ at $m_c = 7$, respectively. Dashed lines indicate $\ln (\tau_{\alpha} / \tau_o) = \Delta G_o z(T) / k_{\mathrm{B}} T$, where $\tau_o$, $L_A$, and $\Delta G_o$ are explained in the text. Panels (c--f) show the variations with $m_c$ and $M$ of the activation enthalpy $\Delta H_o$ and entropy $\Delta S_o$ over a range of fixed $M$ and $m_c$, respectively. The results for linear polymer melts are included as a reference in panels (d) and (f).}
	\label{Fig_String}
\end{figure*}

To describe the dynamics below the onset temperature $T_A$, where relaxation starts to become overtly non-Arrhenius, the string model assumes that the activation free energy $\Delta G(T)$ for structural relaxation is proportional to a factor $z(T)$ given by the average string length $L(T)$ at a given $T$ normalized by its value $L_A$ at $T_A$, $z(T) = L(T)/L_A$. (A detailed discussion of $T_A$ for the present knotted ring polymer melts can be found in our previous paper. \cite{Glass_2024_57_6875}) This assumption yields a temperature-dependent activation free energy $\Delta G(T) = \Delta G_o z(T)$, where $\Delta G_o$ is the high-temperature activation free energy, taking the form of $\Delta G_o = \Delta H_o - T\Delta S_o$ with the enthalpic $\Delta H_o$ and entropic $\Delta S_o$ contributions. The string model then yields the following expression for $\tau_{\alpha}$,
\begin{eqnarray}
	\label{Eq_String}
	\tau_{\alpha} = \tau_{o} \exp\left[ \frac{\Delta G_o}{k_{\mathrm{B}}T} z(T) \right], \ z(T) = L(T)/L_A
\end{eqnarray}
The parameter $\tau_{o}$ can be eliminated by using the value of $\tau_{\alpha}$ at $T_A$, and $\Delta H_o$ can be determined from the Arrhenius fit of the $T$ dependence of $\tau_{\alpha}$ in the high-temperature regime, as discussed in our previous paper. \cite{Glass_2024_57_6875} These considerations result in the following equation, with $\Delta S_o$ being the only fitting parameter,
\begin{eqnarray}
	\label{Eq_L}
	\tau_{\alpha} = \tau_{\alpha} (T_A)\exp\left[\frac{\Delta H_o - T\Delta S_o}{k_{\mathrm{B}}T} z(T) - \frac{\Delta H_o - T_A\Delta S_o}{k_{\mathrm{B}}T_A}\right]
\end{eqnarray}
We use the above equation to test the validity of the string model. As illustrated in Figure \ref{Fig_String}a, b, all our simulation data, spanning a range of $T$, $m_c$, and $M$, can be satisfactorily described by eq \ref{Eq_L}. This finding implies that the pronounced variations of $\tau_\alpha$ with $m_c$ and $M$ can indeed be understood based on the extent of stringlike cooperative motion.

We summarize the fitted activation parameters $\Delta H_o$ and $\Delta S_o$ in Figure \ref{Fig_String}c--f. In previous works based on other polymeric systems, \cite{Unifying_2015_142_234907, Interfacial_2014_5_4163, Quantitative_2015_112_2966, Parallel_2023_56_4929} $\Delta H_o$ and $\Delta S_o$ were often found to vary proportionally, an effect known as the entropy-enthalpy compensation. \cite{Mass_2015_143_144905, Enthalpyentropy_2020_11_3977} Our previous simulation works also showed strong positive correlations in polymer melts with variable cohesive energy \cite{Role_2020_53_9678} and pressure, \cite{Investigation_2020_53_6828} but a negative correlation observed in polymer melts with varying chain stiffness is evidently not ``universal''. \cite{Molecular_2020_53_4796} Our analysis in Figure \ref{Fig_String} indicates that there is a correlation between $\Delta H_o$ and $\Delta S_o$ in knotted ring polymer melts, despite the absence of a universal relationship.

\subsection{\label{Sec_Localization}Localization Model Description of Glass Formation}

After showing that the segmental relaxation in knotted ring polymer melts can be understood based on the stringlike cooperative motion, we examine whether or not other theoretical frameworks provide a reasonable description of the same phenomenon. More specifically, this section considers the localization model of glass formation, another framework that has attracted considerable interest recently. \cite{Localization_2016_2016_054048, Generalized_2012_8_11455, Quantitative_2015_112_2966} The localization model links the structural relaxation time $\tau_\alpha$ to the Debye-Waller parameter $\langle u^2 \rangle$. This quantity defines a measure of local stiffness and can be determined by the mean square displacement (MSD) on a caging time scale on the order of a picosecond. \cite{Universal_2008_4_42, Universal_2009_131_224517, Universal_2011_357_298, Scaling_2011_115_14046} The definition of the MSD can be found in Section S1 of the Supporting Information, along with typical examples. $\langle u^2 \rangle$ can be regarded as a measure of the dynamical free volume and thus provides insights into local mobility gradients that are not evident from a structural perspective. \cite{What_2002_89_125501} Moreover, $\langle u^2 \rangle$ has been linked to atomic-scale stiffness and the overall shear modulus of the entire material, offering insight into local fluctuations and deformation of glass-forming materials. \cite{Initiation_2021_155_204504} We discuss this point in further detail in Section \ref{Sec_Elastic}. Since the non-ergodicity parameter $f_{s, q^*}$ also provides a measure of material stiffness, \cite{Parallel_2023_56_4929} the relationship between $f_{s, q^*}$ and $\langle u^2 \rangle$ in knotted ring polymer melts is briefly discussed in the Section S2 of the Supporting Information.

\begin{figure*}[htb!]
	\centering
	\includegraphics[angle=0, width=0.8\textwidth]{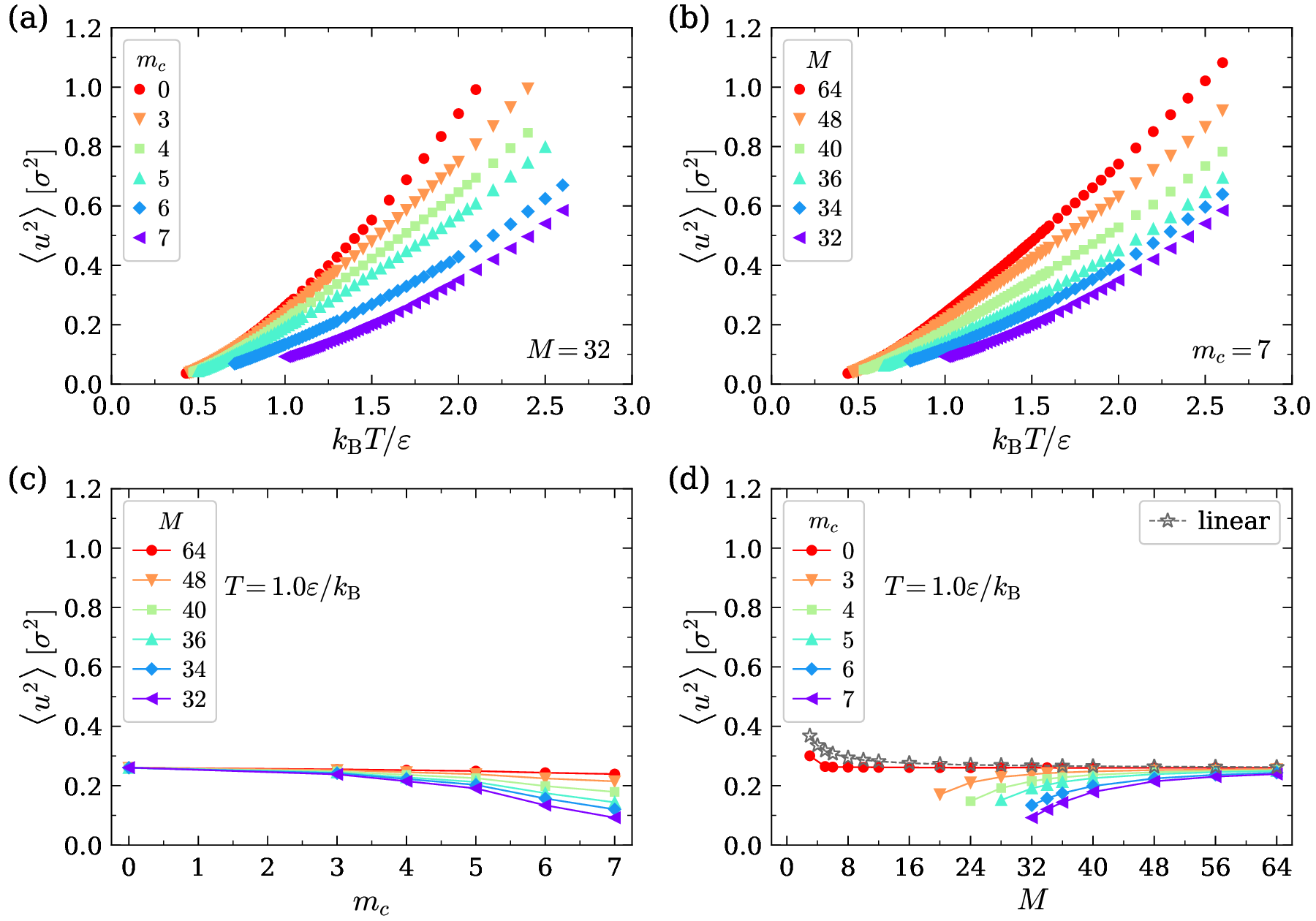}
	\caption{Debye-Waller parameter $\langle u^2 \rangle$ in variation with temperature $T$, knot complexity $m_c$, and molecular mass $M$. Panels (a) and (b) show $\langle u^2 \rangle$ versus $ k_{\mathrm{B}} T / \varepsilon$ for a range of $m_c$ at $M = 32$ and for a range of $M$ at $m_c = 7$, respectively. Panels (c) and (d) show the variations with $m_c$ and $M$ of $\langle u^2 \rangle$ over a range of fixed $M$ and $m_c$ at $T = 1.0 \varepsilon / k_{\mathrm{B}}$, respectively. The results for linear polymer melts are included as a reference in panel (d).}
	\label{Fig_DWF}
\end{figure*}

We first discuss the variations of $\langle u^2 \rangle$ with $T$, $m_c$, and $M$ in knotted ring polymer melts, as given in Figure \ref{Fig_DWF}. Again, we see that both $m_c$ and $M$ can substantially influence the magnitude of $\langle u^2 \rangle$. As shown in Figure \ref{Fig_DWF}c, the dependence of $\langle u^2 \rangle$ on $m_c$ is generally weak, except for a pronounced decrease when $m_c$ is raised to $5$ in the low $M$ regime. In contrast, when $M$ is large, the variation of $\langle u^2 \rangle$ with $m_c$ remains weak across the entire range of $m_c$ considered. Furthermore, Figure \ref{Fig_DWF}d demonstrates that $\langle u^2 \rangle$ gradually decreases as $M$ decreases, a trend that deviates from those for linear and unknotted ring polymer melts. Overall, when $T$, $m_c$, and $M$ are varied, $\langle u^2 \rangle$ exhibits similar trends to those discussed above for $\tau_{\alpha}$ and $L$. It is then of interest to explore the possibility of a quantitative connection between $\tau_{\alpha}$ and $\langle u^2 \rangle$.

The original formulation of the localization model proposes that $\tau_\alpha$ should be related to $\langle u^2 \rangle$ by the following relation,
\begin{equation}
	\label{Eq_Local}
	\tau_{\alpha} = \tau_u \exp \left[ (u_0^2 / \langle u^2 \rangle)^{\alpha/2} \right]
\end{equation}
where the parameters $\tau_u$ and $u_0^2$ can be specified in terms of the simulation data for $\tau_\alpha$ and $\langle u^2 \rangle$ at $T_A$, namely, $\tau_A = \tau_\alpha (T_A)$ and $u_A^2 = \langle u^2(T_A) \rangle$, and $\alpha$ is an adjustable parameter in the model discussed below. The localization model then adopts the remarkably simple relation,
\begin{equation}
	\label{Eq_Local2}
	\tau_{\alpha} = \tau_A \exp \left[ (u_A^2 / \langle u^2 \rangle)^{\alpha/2} - 1\right]
\end{equation}

\begin{figure*}[htb!]
	\centering
	\includegraphics[angle=0, width=0.8\textwidth]{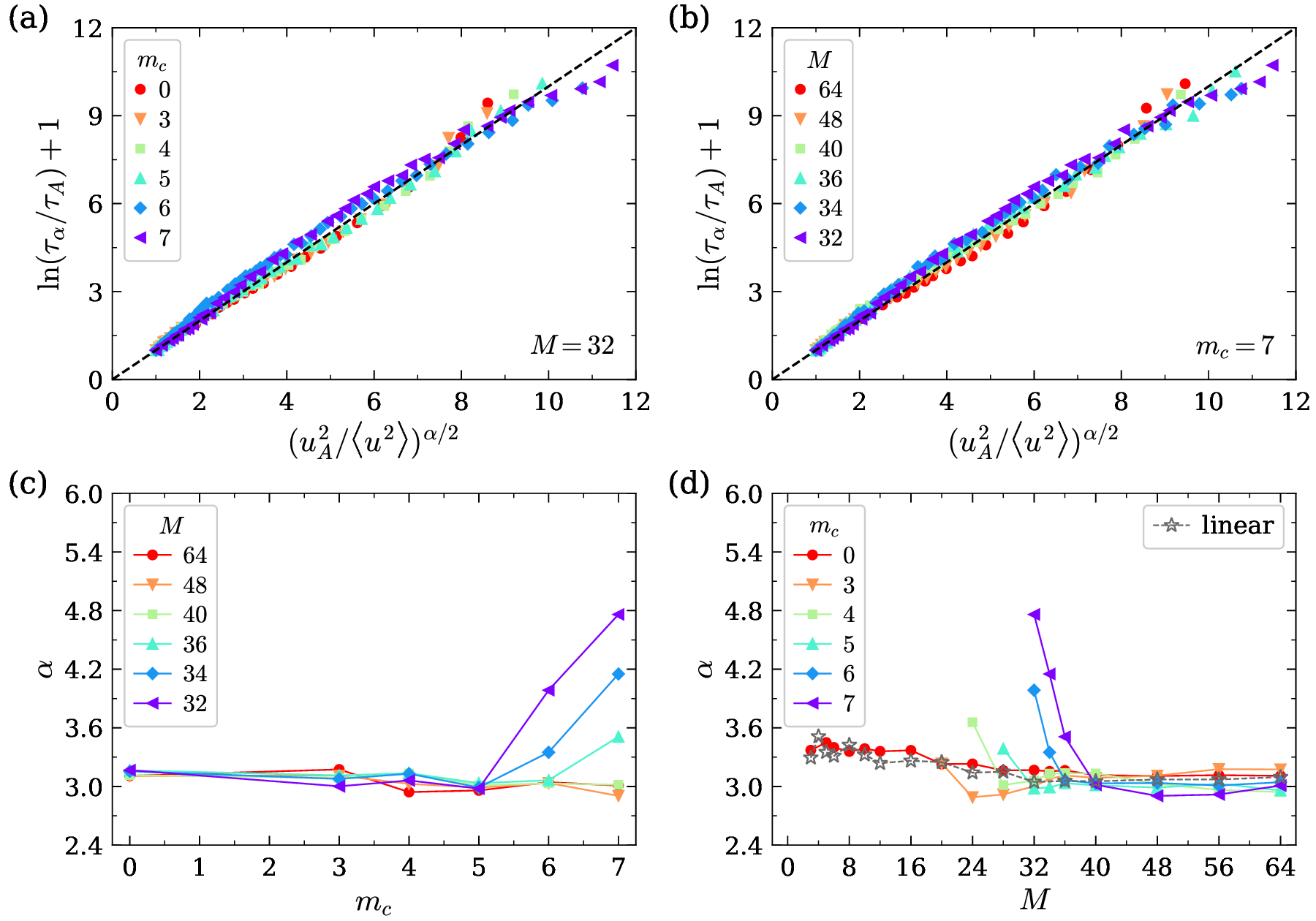}
	\caption{Localization model description of glass formation. Panels (a) and (b) show $\ln (\tau_{\alpha} / \tau_A) + 1$ versus $(u_A^2 / \langle u^2 \rangle)^{\alpha/2}$ for a range of $m_c$ at $M = 32$ and for a range of $M$ at $m_c = 7$, respectively. Dashed lines indicate $\ln (\tau_{\alpha} / \tau_A) + 1 = (u_A^2 / \langle u^2 \rangle)^{\alpha/2}$, where $\tau_A$, $u_A^2$, and $\alpha$ are explained in the text. Panels (c) and (d) show the variations with $m_c$ and $M$ of $\alpha$ over a range of fixed $M$ and $m_c$, respectively. The results for linear polymer melts are included as a reference in panel (d).}
	\label{Fig_Local}
\end{figure*}

A test of the localization model of glass formation is shown in Figure \ref{Fig_Local}a, b, where we find that this model provides a satisfactory description of our simulation data for knotted ring polymer melts. Figure \ref{Fig_Local}c shows the variation of $\alpha$ with $m_c$. It can be seen that $\alpha$ remains relatively constant for large values of $M$ as $m_c$ changes, while for smaller values of $M$, $\alpha$ initially remains unchanged in magnitude with respect to $m_c$, but $\alpha$ exhibits a significant increase at $m_c = 5$ to values that evidently exceed $3$. Figure \ref{Fig_Local}d illustrates the variation of $\alpha$ with $M$. For small $m_c$, $\alpha$ for knotted ring polymer melts depends weakly on $M$, similar to that of linear and unknotted ring polymers, but $\alpha$ increases with decreasing $M$ when $m_c$ is large.

Remarkably, the localization model involves no free parameter if the exponent $\alpha$ is taken to be the spatial dimensionality $d$ so that $\langle u^2 \rangle^{\alpha/2}$ has the units of volume. In the original investigation of the localization model computational by Starr and coworkers, \cite{Quantitative_2015_112_2966, Explaining_2021_127_277802, Relating_2022_157_131101} the dependence of $\tau_\alpha$ on $\langle u^2 \rangle$ for model polymer melts indicated an exponent consistent with $\alpha = 3$. Later extensive simulations of the Kob-Andersen model system under a wide range of densities and pressures as well as a Zr-Cu metallic glass also indicated consistency with $\alpha = 3$. \cite{Understanding_2024_128_10999, Localization_2016_2016_054048} Additionally, it was found that the interfacial dynamics of both crystalline and metallic glass-forming materials can be well described by the localization model with $\alpha = 3$. \cite{Localization_2021_44_33, Dynamics_2022_157_114505} Despite the fact that $\alpha$ has often been found to be near $3$, there have also been reports of apparent $\alpha$ values differing appreciably from $3$. \cite{Tuning_2015_4_1134, Generalized_2012_8_11455, Roles_2015_53_1458} In our previous work on the Kob-Andersen model system, \cite{Understanding_2024_128_10999} we also found that $\alpha$ could vary over an appreciable range until we realized that the choice of time at which $\langle u^2 \rangle$ is defined can exert an appreciable influence on $\alpha$ and that the uncertainty in the estimation of $T_A$ can also influence the apparent value of $\alpha$. By applying a physically robust criterion, where $t_{\mathrm{onset}}$ corresponds to the timescale at which the fast $\beta$- and $\alpha$-relaxation processes merge, we confirm that while the choice of $t_{\mathrm{onset}}$ can shift the absolute magnitude of $\alpha$, the qualitative trend of $\alpha$ increasing above $3$ in the ``tight knot'' limit remains robust. This observation aligns with recent theoretical reformulations of the localization model, \cite{Dynamical_2024_20_9140} suggesting that $\alpha$ is not a universal constant, but rather a variable reflecting the degree of anharmonicity of intermolecular interactions. To quantitatively test this link between the non-universal behavior of $\alpha$ and the nature of particle localization in knotted polymers, we next analyze the degree of non-linearity in the $T$ dependence of $\langle u^2 \rangle$, which serves as a practical measure of intermolecular anharmonicity.

To illustrate this effect, we first calculated $\langle u^2 \rangle$ for our polymer melts based on the assumption that the characteristic time scale for defining $\langle u^2 \rangle$ was equal to $1 \tau$. This reduced time corresponds to a time scale on the order of a picosecond in laboratory units, which is a convention used in previous simulations of polymer liquids \cite{Stringlike_2018_148_104508} and this time scale serves as a good approximation of the $\alpha$-relaxation time at $T_A$ for our model polymer melts. Yuan et al. \cite{Influence_2024_128_10999, Understanding_2024_128_10999} have strongly advocated that $\tau_\alpha(T_A)$ is the appropriate definition of the caging onset time scale defining $\langle u^2 \rangle$ based on a thorough analysis of the Kob-Andersen model system. In Section S3 of the Supporting Information, we also show a test of the localization model based on the alternative assumption of a caging onset time, $t_{\mathrm{onset}} = 2 \tau$. Indeed, this larger choice of $t_{\mathrm{onset}}$ leads to values of $\alpha$ closer to $2.5$ for linear and unknotted polymer melts. Our estimates of $\alpha$ based on the criterion $t_{\mathrm{onset}} = 1 \tau$, which we think is a more suitable criterion, are shown in Figure \ref{Fig_Local}c, d.

Our results in Figure \ref{Fig_Local}c, d reveal an unexpected trend in $\alpha$, with increasing $m_c$ and in the limit of ``tight'' knots having low $M$ where there seems to be a real trend for $\alpha$ to approach values larger than $3$. (The same qualitative increase is also apparent in $\alpha$ estimates based on $t_{\mathrm{onset}} = 2 \tau$.) This trend in $\alpha$ looks convincing, so the generality of $\alpha = 3$ is clearly suspect. The source of this variability of $\alpha$ then requires a discussion based on existing formulations of the localization model.

Simmons et al. \cite{Generalized_2012_8_11455} heuristically argued in an early formulation of the localization model that the volume dynamically explored by chain segments in complex fluids, such as polymer melts, could lead to anisotropic, thread-like dynamical free volume regions instead of compact quasi-spherical regions. Based on this physical model, it was then suggested that this effect could plausibly result in values of the effective dimension parameter $\alpha$ being less than the spatial dimensionality $d$, and that anisotropy of the dynamical free volume regions could naturally explain some potential variability of $\alpha$ with the fluid type. Unfortunately, the physically appealing geometrical model of the dynamical free volume ``structure'' cannot explain values of $\alpha$ larger than $3$ in three-dimensional simulations, as reported in some previous simulations noted above and in Figure \ref{Fig_Local}, depicting our simulations for ring polymer melts of variable $m_c$ and $M$. Evidently, this model of $\alpha$ variability is simply not adequate to explain observed trends in apparent $\alpha$ values. As noted previously, Douglas et al. \cite{Dynamical_2024_20_9140} reformulated the localization model based on a general, but rather abstract dynamical systems perspective. The magnitude of $\alpha$ is predicted to vary in this formulation, but its value reflects the degree of anharmonicity of the intermolecular interaction, making it difficult to estimate the precise value of $\alpha$ for any particular molecular fluid. On the other hand, it is possible to gain qualitative insight into the variation of $\alpha$ in fluids through the quantification of the degree of non-linearity in the $T$ dependence of $\langle u^2 \rangle$, a well-known practical measure of the anharmonicity of the intermolecular interaction. \cite{Correlation_2001_87_215901, Stringlike_2018_148_104508} 

\begin{figure*}[htb!]
	\centering
	\includegraphics[angle=0, width=0.8\textwidth]{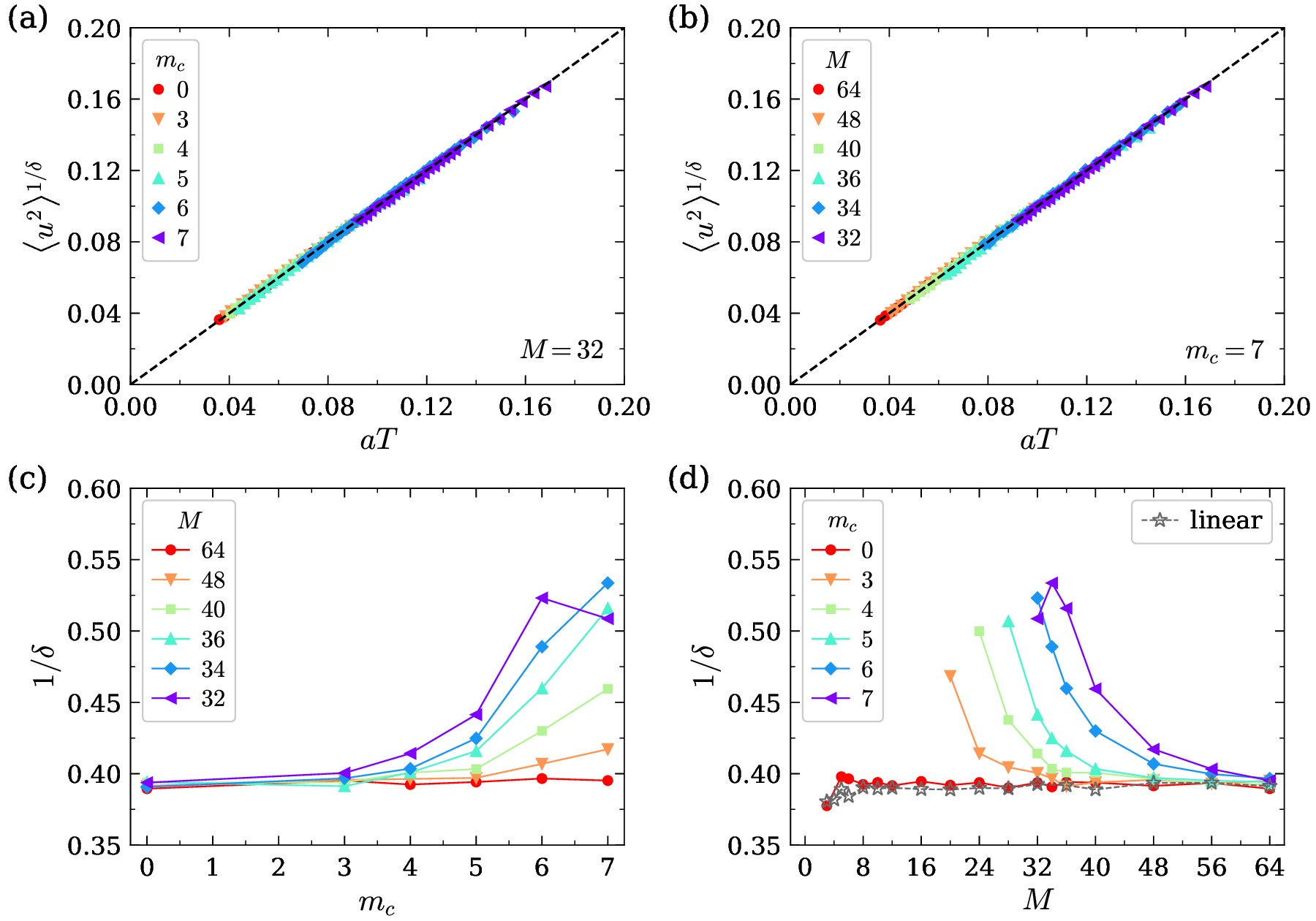}
	\caption{Analysis of the temperature dependence of the Debye-Waller parameter $\langle u^2 \rangle$. Panels (a) and (b) show $\langle u^2 \rangle^{1/\delta}$ versus $a T$ for a range of $m_c$ at $M = 32$ and for a range of $M$ at $m_c = 7$, respectively. Dashed lines indicate $\langle u^2 \rangle^{1/\delta} = aT$, where $\delta$ and $a$ are fitting parameters that depend on $m_c$ and $M$. Panels (c) and (d) show the variations with $m_c$ and $M$ of $1/\delta$ over a range of fixed $M$ and $m_c$, respectively. The results for linear polymer melts are included as a reference in panel (d).}
	\label{Fig_DWF_T}
\end{figure*}

Betancourt et al. \cite{Stringlike_2018_148_104508} introduced a convenient method for quantifying the nonlinear $T$ dependence of $\langle u^2 \rangle$, through the nonlinear transformation, $\langle u^2 \rangle^{1/\delta} \sim T$, where $\delta$ is an ``anharmonicity parameter'', given that a linear $T$ scaling of $\langle u^2 \rangle$ is characteristic of harmonically localized particles. The use of such non-linear scaling of $\langle u^2 \rangle$ to characterize material anharmonicity has been established in early studies of various condensed systems. \cite{Empirical_1977_48_494, EXAFS_1984_52_511, Anharmonic_1997_56_43} Notably, this method can apparently be applied readily to essentially any fluid. Accordingly, we analyze the influence of $m_c$ and $M$ on the degree of anharmonicity based on such a method which has not been applied to any specific purpose previously. Figure \ref{Fig_DWF_T}a, b indicates that a linear relation between $\langle u^2 \rangle^{1/\delta}$ and $T$ generally holds quite well in knotted ring polymer melts. Figure \ref{Fig_DWF_T}c, d summarizes the fitted results for $1/\delta$ as a function of $m_c$ and $M$. As can be seen, the degree of anharmonicity $\delta$ increases with increasing $m_c$ for small $M$ and with decreasing $M$ for large $m_c$, exhibiting the same general trends for $\alpha$ observed in Figure \ref{Fig_Local}c, d. In the case of $M = 32$ and $m_c = 7$, $1/\delta$ exhibits a nonmonotonic variation, which might arise from non-equilibrium effects associated with knot rigidification in these limits.

Although it would be truly remarkable if the structural relaxation time $\tau_{\alpha}$ can be predicted based on $\langle u^2 \rangle$ estimates for all fluids without any free parameters, as followed from the localization model when $\alpha$ equals $3$, there is apparently some variability in the exponent $\alpha$ for some systems that can spoil this ``universality''. The model of knotted ring polymers has allowed us to probe this non-universality and gain insight into its cause.

\subsection{\label{Sec_Shoving}Shoving Model Description of Glass Formation}

In addition to the string and localization models, there are alternative models of liquid dynamics characterized by fundamentally different quantities. This section focuses on the shoving model, which attributes the dramatic slowdown of structural relaxation and diffusion in glass‐forming materials to the increase in shear modulus or other measures of material stiffness. The glassy plateau shear modulus $G_p$ is often measured as the high-frequency shear modulus in viscoelastic materials since the measurements of material stiffness have been suggested to determine the temperature-dependent activation free energy $\Delta G(t)$ of glass-forming liquids. \cite{Temperature_1965_43_139, Relationship_2013_138_12A541, String_2014_140_204509, Polymer_2021_54_3001} Specifically, $G_p$ is estimated from the stress autocorrelation function $G(t)$ at the caging onset time, i.e., $G_p \equiv G(t = 1\tau)$, as in the case of $\langle u^2 \rangle$ and $f_{s, q^*}$, while the infinite-frequency modulus is given by $G_\infty = G(t \to 0^+)$. The definition of $G(t)$ can be found in Section S1 of the Supporting Information, along with typical examples.

\begin{figure*}[htb!]
	\centering
	\includegraphics[angle=0, width=0.8\textwidth]{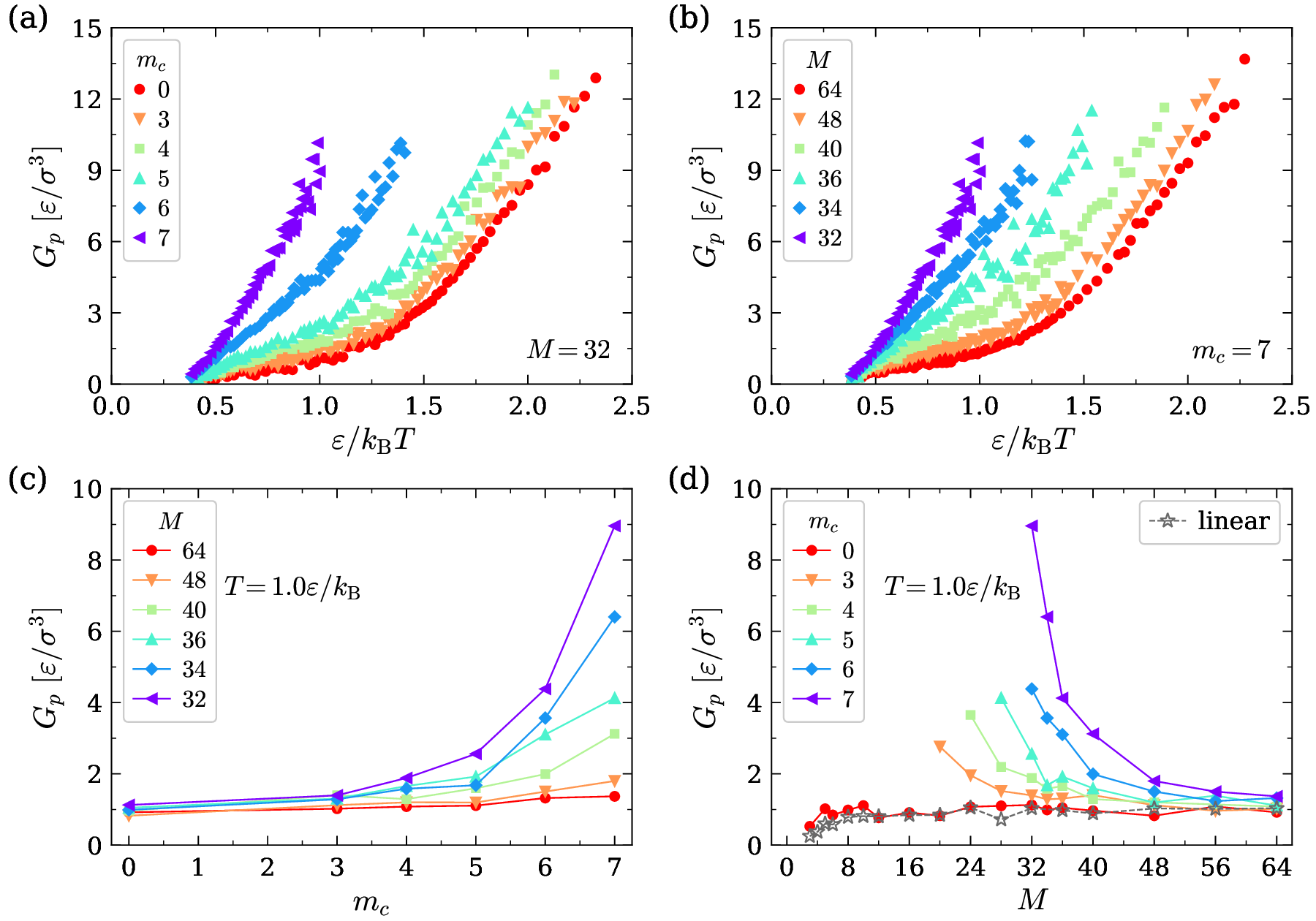}
	\caption{Glassy plateau shear modulus $G_p$ in variation with temperature $T$, knot complexity $m_c$, and molecular mass $M$. Panels (a) and (b) show $G_p$ versus $\varepsilon / k_{\mathrm{B}} T$ for a range of $m_c$ at $M = 32$ and for a range of $M$ at $m_c = 7$, respectively. Panels (c) and (d) show the variations with $m_c$ and $M$ of $G_p$ over a range of fixed $M$ and $m_c$ at $T = 1.0 \varepsilon / k_{\mathrm{B}}$, respectively. The results for linear polymer melts are included as a reference in panel (d).}
	\label{Fig_Gp}
\end{figure*}

Figure \ref{Fig_Gp} summarizes the representative results for the variations of $G_p$ with $T$, $m_c$, and $M$. It is clear that $G_p$ follows analogous patterns to those observed for $\tau_{\alpha}$, $L$, and $\langle u^2 \rangle$. The relationship between $G_p$ and $T$ is illustrated in Figure \ref{Fig_Gp}a for varying $m_c$ at $M = 32$, and in Figure \ref{Fig_Gp}b for varying $M$ at $m_c = 7$. These findings indicate that both $m_c$ and $M$ can markedly affect $G_p$. To investigate the effects of $m_c$ and $M$ on $G_p$ at a constant $T$, Figure \ref{Fig_Gp}c reveals that $G_p$ initially exhibits a weak dependence on $m_c$, followed by a large increase around $m_c = 5$ in the low $M$ regime. At sufficiently high $M$, the influence of knotting on $G_p$ remains minimal across the entire range of $m_c$ examined in this study, with results aligning well with those of linear and unknotted ring polymers. Note that our simulation data for $G_p$ are more scattered than the other data such as $\tau_{\alpha}$, $L$, and $\langle u^2 \rangle$, because the stress autocorrelation function is a property of the overall system that cannot be averaged over individual particles.

We now examine the extent to which structural relaxation can be described by the shoving model. A key prediction of this model is that structural relaxation is activated by an energy barrier $\Delta E$, which is proportional to an elastic modulus explicitly identified as the instantaneous shear modulus $G_{\mathrm{inst}}$ due to the cage breaking,
\begin{equation}
	\label{Eq_Elastic}
	\tau_{\alpha} = \tau_G \exp(\Delta E / k_{\mathrm{B}} T),\ \Delta E = G_{\mathrm{inst}} V_c^*
\end{equation}
where $\tau_G$ is a prefactor and $V_c^*$ is an empirical characteristic volume, which is assumed to be on the order of a segment volume and independent of $T$. Puosi and Leporini \cite{Communication_2012_136_041104} first emphasized that $G_{\mathrm{inst}}$ should not be identified with the infinite frequency shear modulus, $G_{\infty}$, which has a well-known thermodynamic definition. Instead, $G_{\infty}$ should be associated with a glassy plateau modulus $G_p$, corresponding to $G(t)$ after an initial fast $\beta$ stress relaxation process, as discussed above. Since the fast $\beta$-relaxation time is generally on the order of $0.1$ to $1$ ps for both molecular and atomic glass-forming liquids, \cite{Light_1997_9_10079, Fast_1997_126_159, Stringlike_2018_148_104508} our definition of $G_p$ at $t = 1 \tau$ is consistent with this picture. The qualitative difference between $G_p$ and $G_{\infty}$ has been noted previously. \cite{Predictive_2020_6_eaaz0777, Communication_2012_136_041104} Dyre and Wang \cite{instantaneous_2012_136_224108} accordingly modified the shoving model by replacing $G_{\infty}$ with $G_p$.

\begin{figure*}[htb!]
	\centering
	\includegraphics[angle=0, width=0.8\textwidth]{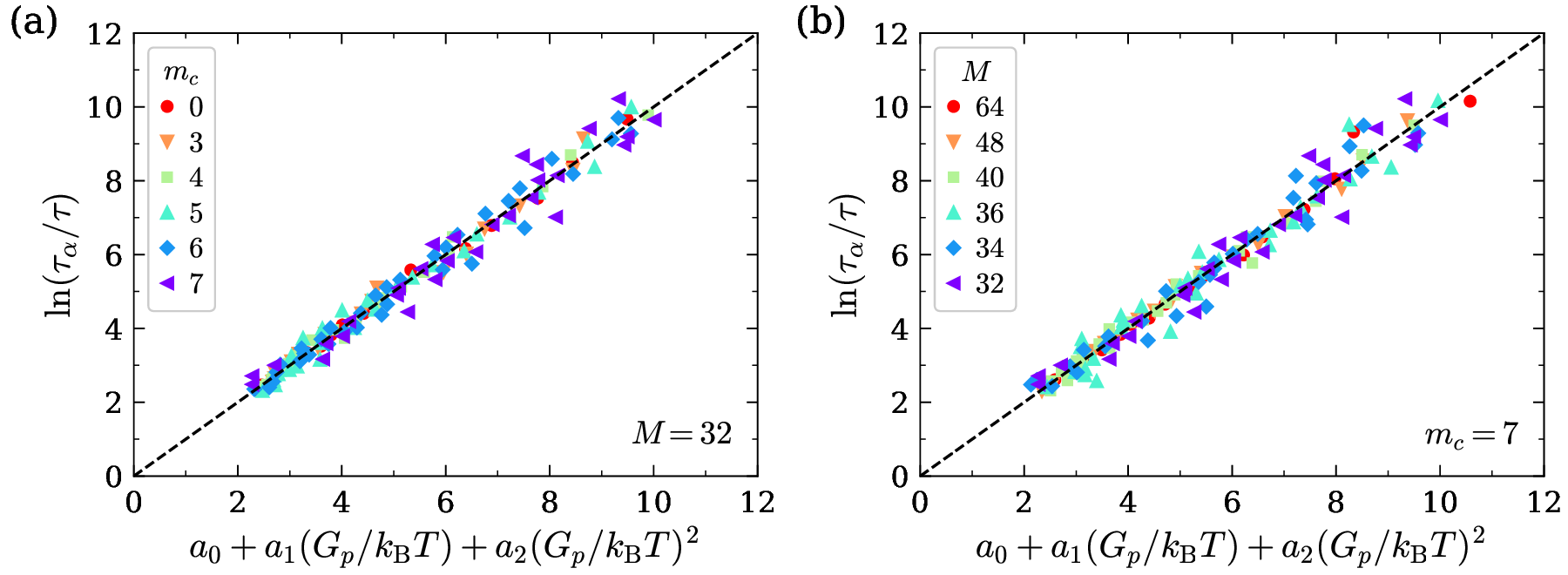}
	\caption{Shoving model description of glass formation. Panels (a) and (b) show $\ln (\tau_{\alpha} / \tau)$ versus $a_0 + a_1(G_p / k_{\mathrm{B}}T) + a_2(G_p / k_{\mathrm{B}}T)^2$ for a range of $m_c$ at $M = 32$ and for a range of $M$ at $m_c = 7$, respectively. $a_0$, $a_1$, and $a_2$ are fitting parameters that depend on $m_c$ and $M$. Lines indicate $\ln (\tau_{\alpha} / \tau) = a_0 + a_1(G_p / k_{\mathrm{B}}T) + a_2(G_p / k_{\mathrm{B}}T)^2$.}
	\label{Fig_Shoving}
\end{figure*}

Puosi and Leporini \cite{Communication_2012_136_041104} further proposed a specific quadratic relationship between $\tau_{\alpha}$ and $G_p$, which was verified for coarse-grained polymer melts and small-molecule liquids,
\begin{equation}
	\label{Eq_Quadratic}
	\ln (\tau_{\alpha}) = a_0 + a_1 (G_p / k_{\mathrm{B}} T) + a_2 (G_p / k_{\mathrm{B}} T)^2
\end{equation}
where $a_0$, $a_1$, and $a_2$ are adjustable constants. Our recent work also showed that the same functional form holds in the Kob-Andersen model system over a wide range of densities and pressures. \cite{Understanding_2024_128_10999} Here, we test such a relation for knotted ring polymer melts with different $m_c$ and $M$. As can be seen in Figure \ref{Fig_Shoving}, eq \ref{Eq_Quadratic} holds reasonably well, although the simulation data are scattered when the knots are tight, due to the reason explained above.

\subsection{\label{Sec_Interrelation}Property Interrelations}

The near equivalence of independent approaches to describing the dynamics of glass-forming liquids, each centering around different fluid properties, has interesting implications. In particular, this finding implies that the properties emphasized by these approaches might be closely interrelated. In this section, we show that it is indeed the case in knotted ring polymer melts.

\begin{figure*}[htb!]
	\centering
	\includegraphics[angle=0, width=0.8\textwidth]{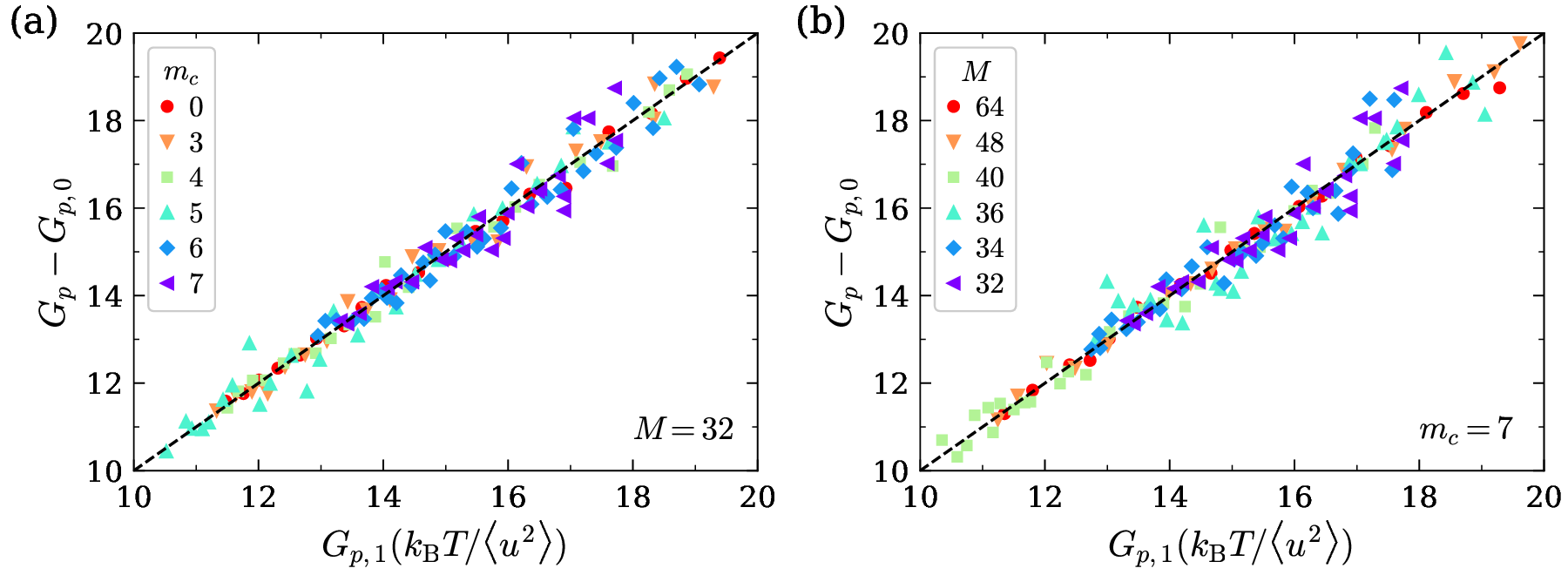}
	\caption{Correlation between the Debye-Waller parameter $\langle u^2 \rangle$ and the glassy plateau shear modulus $G_p$. Panels (a) and (b) show $G_p - G_{p,0}$ versus $G_{p,1}(k_{\mathrm{B}} T / \langle u^2 \rangle)$ for a range of $m_c$ at $M = 32$ and for a range of $M$ at $m_c = 7$, respectively. $G_{p,0}$ and $G_{p,1}$ are fitting parameters that depend on $m_c$ and $M$. Lines indicate $G_p - G_{p,0} = G_{p,1}(k_{\mathrm{B}} T / \langle u^2 \rangle)$.}
	\label{Fig_Gp_DWF}
\end{figure*}

One nontrivial property interrelation is that between the high frequency modulus $G_p$ of the entire material and $\langle u^2 \rangle$, a molecular scale measure of material stiffness. In particular, previous works have established the nontrivial relation, \cite{kinetic_2015_38_87, Communication_2012_136_041104} 
\begin{equation}
	\label{Eq_GpDWF}
	G_p \approx G_{p,0} + G_{p,1} (k_{\mathrm{B}} T / \langle u^2 \rangle)
\end{equation}
where $G_{p,0}$ and $G_{p,1}$ are material-dependent constants. This remarkable relation has been confirmed for many materials, \cite{Understanding_2022_157_064901, Competing_2022_55_9990, kinetic_2015_38_87, Initiation_2021_155_204504, Parallel_2023_56_4929} and in Figure \ref{Fig_Gp_DWF}, we show that eq \ref{Eq_GpDWF} is also applicable to knotted ring polymer melts, where the constants $G_{p,0}$ and $G_{p,1}$ depend on $m_c$ and $M$. This relation also enables an investigation of the \textit{local} stiffness of a material based on the distribution of $\langle u^2 \rangle$, namely, the elastic heterogeneity, as discussed in Section \ref{Sec_Elastic}. It has been established that the bulk modulus, \cite{Equation_2021_54_3247, Understanding_2022_157_064901, Competing_2022_55_9990, Influence_2023_56_3873, Influence_2023_56_7636} central to some models of the dynamics of condensed materials, \cite{Activation_2021_155_174901} and the configurational entropy $S_c$, \cite{Equation_2021_54_3247, Understanding_2024_128_10999} central to the AG model of glass formation, \cite{Temperature_1965_43_139} can also be related to $\langle u^2 \rangle$ quantitatively.

\begin{figure*}[htb!]
	\centering
	\includegraphics[angle=0, width=0.8\textwidth]{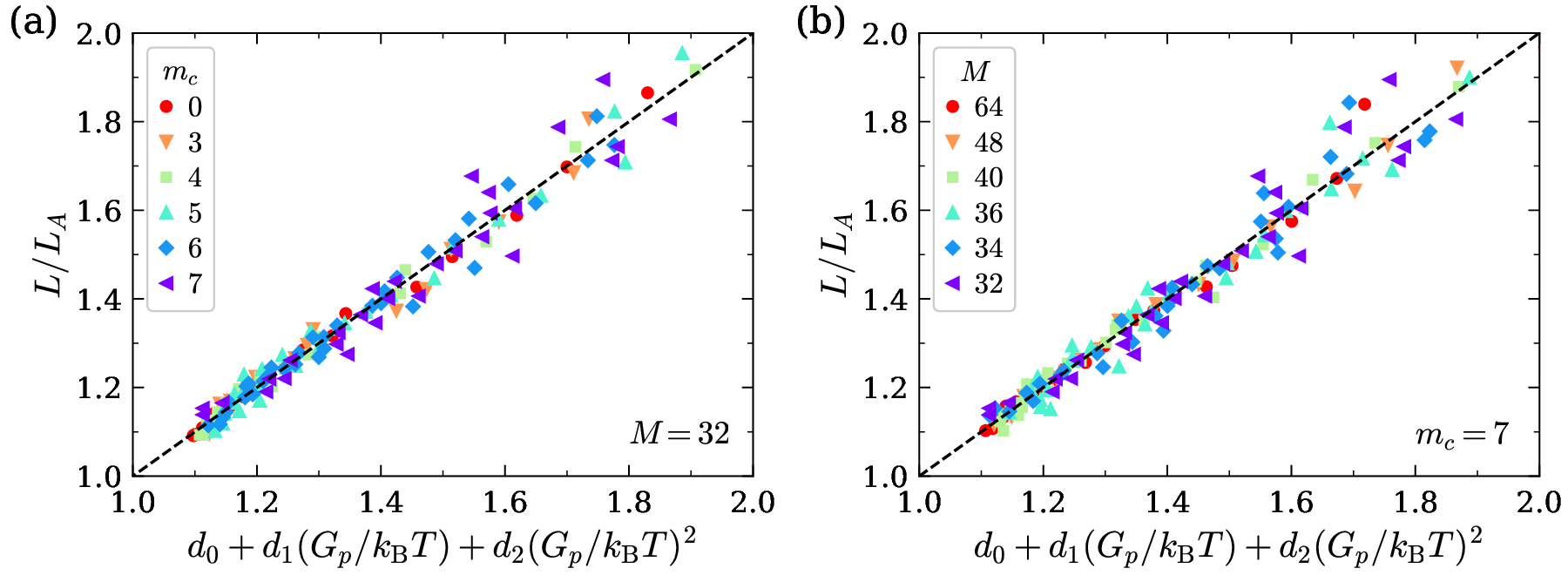}
	\caption{Correlation between the glassy plateau shear modulus $G_p$ and the extent of collective motion $L/L_A$. Panels (a) and (b) show $L/L_A$ versus $d_0 + d_1(G_p / k_{\mathrm{B}}T) + d_2(G_p / k_{\mathrm{B}}T)^2$ for a range of $m_c$ at $M = 32$ and for a range of $M$ at $m_c = 7$, respectively. $d_0$, $d_1$, and $d_2$ are fitting parameters that depend on $m_c$ and $M$. Lines indicate $L/L_A = d_0 + d_1(G_p / k_{\mathrm{B}}T) + d_2(G_p / k_{\mathrm{B}}T)^2$.}
	\label{Fig_LGp}
\end{figure*}

Another nontrivial correspondence of particular interest is the relation between cooperative motion and rigidity in knotted polymers, a phenomenon discussed speculatively by de Gennes. \cite{Tight_1984_17_703} Based on the results discussed above for the string and shoving models, we discuss this type of relation in knotted ring polymer melts. In particular, we may expect a direct scaling relationship between the glassy plateau shear modulus $G_p$ and the extent of collective motion $L/L_A$, a relationship considered previously in linear polymer melts. \cite{Parallel_2023_56_4929} Reminiscent of the functional form of eq \ref{Eq_Quadratic}, we find in Figure \ref{Fig_LGp} that a quadratic relation describes the relationship between $L$ and $G_p$ quite well in knotted polymer melts,
\begin{equation}
	\label{Eq_LGp}
	L/L_A = d_0 + d_1 (G_p / k_{\mathrm{B}} T) + d_2 (G_p / k_{\mathrm{B}} T)^2
\end{equation}
where $d_0$, $d_1$, and $d_2$ are fitting parameters. This scaling reflects the parallel emergence of rigidity and collective motion that underlies applicability of both the string and shoving models to the dynamics of glass-forming liquids. This relation between collective motion and stiffness can also be considered by applying the same analysis to individual knotted polymers, which we plan to perform in future work.

\subsection{\label{Sec_Dynamic}Dynamic Heterogeneity}

It is now well established that the dynamics of glass-forming liquids become increasingly spatially heterogeneous upon approaching $T_{\mathrm{g}}$. In Section \ref{Sec_String}, we discuss the role of stringlike cooperative motion in understanding the segmental relaxation of knotted ring polymer melts, which represents one specific characterization of dynamic heterogeneity. Here, we examine dynamic heterogeneity from other complementary perspectives. In particular, we focus on the non-Gaussian parameter $\alpha_{2}(t)$, which characterizes the dynamic heterogeneity associated with mobile particles, because our previous works identified an intriguing relation between this quantity and the relaxation time in the Kob-Andersen model system. \cite{Understanding_2024_128_10999} $\alpha_{2}(t)$ displays a primary peak, denoted as $\alpha_2^*$, at a peak time $t^*$, which is related to the lifetime of the mobile particle clusters. \cite{Relationship_2013_138_12A541, Role_2015_142_164506} The definition of $\alpha_{2}(t)$ can be found in Section S1 of the Supporting Information, along with typical examples.

\begin{figure*}[htb!]
	\centering
	\includegraphics[angle=0, width=0.8\textwidth]{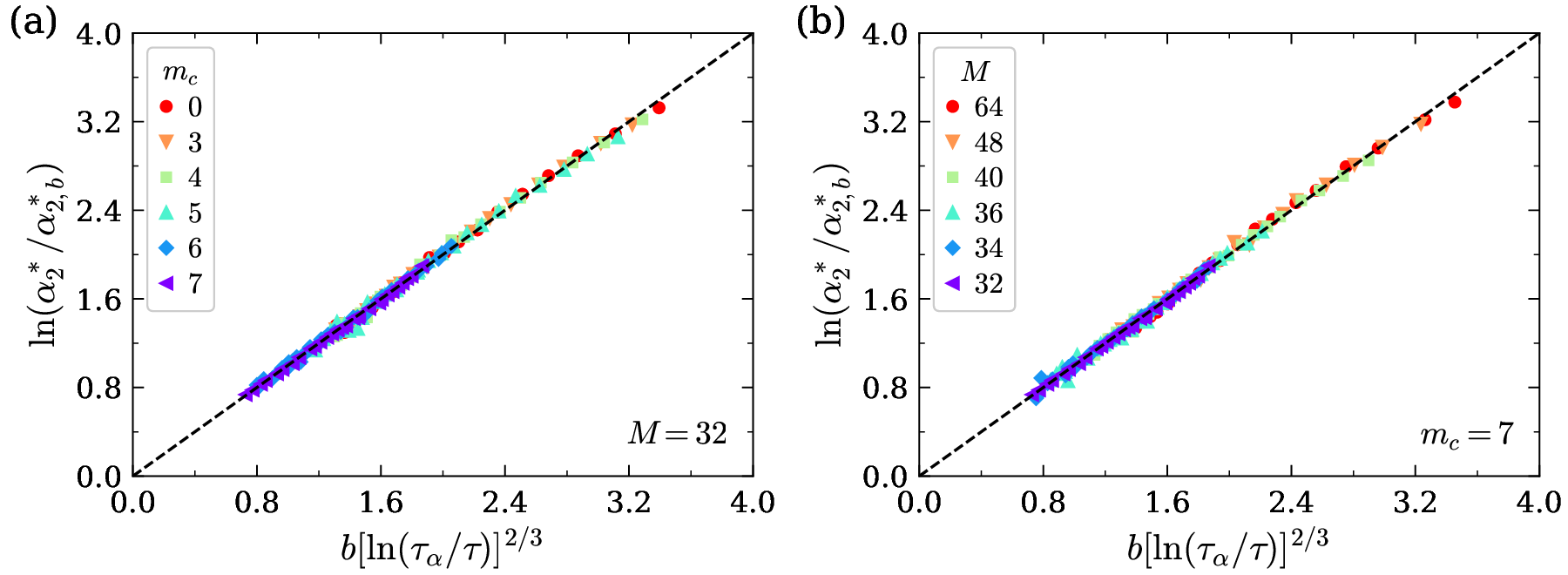}
	\caption{Relation between the structural relaxation time $\tau_\alpha$ and the peak magnitude $\alpha_2^*$ of the non-Gaussian parameter. Panels (a) and (b) show $\ln(\alpha_{2}^* / \alpha_{2,b}^*)$ versus $b \left[ \ln(\tau_a / \tau) \right]^{2/3}$ for a range of $m_c$ at $M = 32$ and for a range of $M$ at $m_c = 7$, respectively. $\alpha_{2,b}^*$ and $b$ are fitting parameters that depend on $m_c$ and $M$. Lines indicate $\ln(\alpha_{2}^* / \alpha_{2,b}^*) = b \left[ \ln(\tau_a / \tau) \right]^{2/3}$.}
	\label{Fig_NGP_Tau}
\end{figure*}

Although the non-Gaussian parameter is commonly interpreted as a measure of the extent of dynamic heterogeneity, its precise physical meaning remains somewhat obscure, beyond the observation that this peak occurs at a characteristic time $t^*$ at which the size of the mobile particle clusters peaks. \cite{Relationship_2013_138_12A541} A potential approach provides some insight into how the magnitude of ``dynamic heterogeneity'', as quantified by $\alpha_2^*$, relates to the magnitude of the structural relaxation time $\tau_\alpha$, as gleaned from the striking correlative relation, \cite{Understanding_2024_128_10999}
\begin{equation}
	\label{Eq_NGPTau}
	\ln(\alpha_{2}^* / \alpha_{2,b}^*) = b \left[\ln(\tau_a / \tau)\right]^{2/3}
\end{equation}
where $\alpha_{2,b}^*$ and $b$ are fitting parameters, as identified in a previous study. \cite{Understanding_2024_128_10999} We take this opportunity to test the generality of this relation for knotted ring polymer melts with different $m_c$ and $M$. As can be seen in Figure \ref{Fig_NGP_Tau}, the relation given by eq \ref{Eq_Quadratic} also holds quite well. It is emphasized that there is currently no theoretical understanding of this striking relationship between $\alpha_2^*$ and $\tau_\alpha$. We develop this expression further by combining it with the string model of glass formation, as discussed in Section \ref{Sec_String}.

\begin{figure*}[htb!]
	\centering
	\includegraphics[angle=0, width=0.8\textwidth]{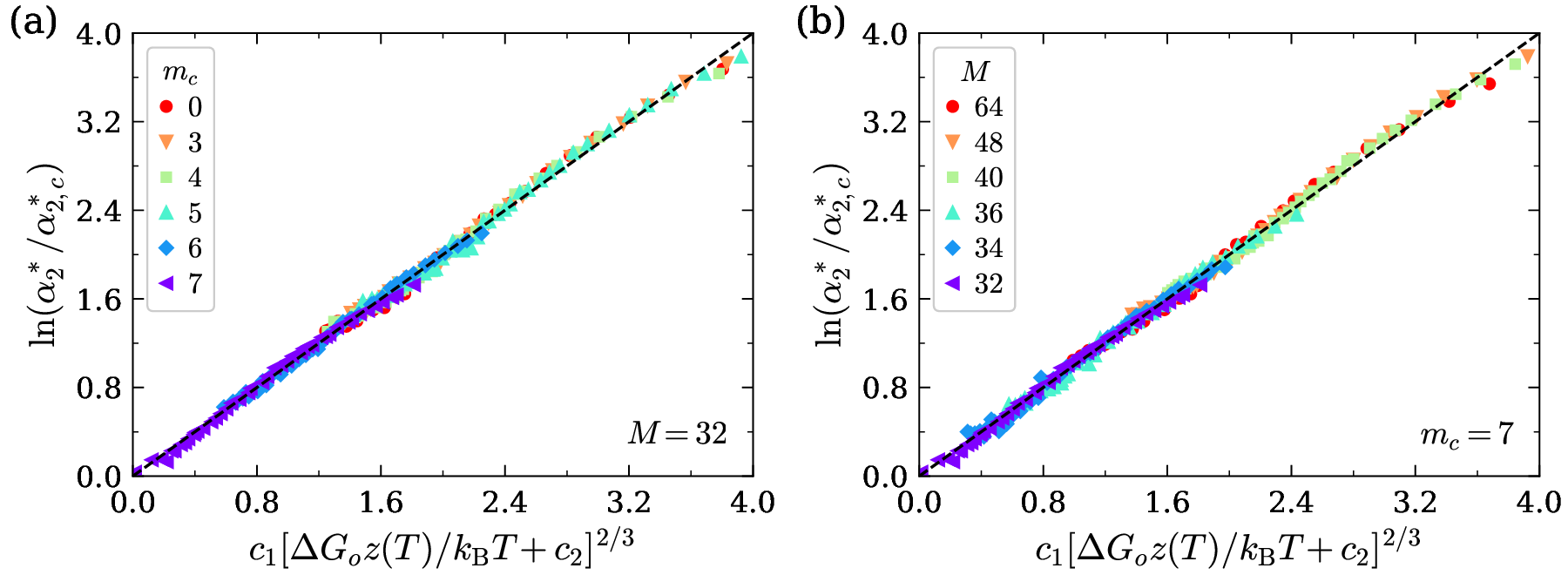}
	\caption{\label{Fig_NGP_String}Relation between the extent of stringlike collective motion $z(T)$ and the peak magnitude $\alpha_2^*$ of the non-Gaussian parameter. Panels (a) and (b) show $\ln(\alpha_{2}^*/\alpha_{2,c}^*)$ versus $c_1 \left[ \Delta G_oz(T)/k_{\mathrm{B}}T +c_2 \right]^{2/3}$ for a range of $m_c$ at $M = 32$ and for a range of $M$ at $m_c = 7$, respectively. $\alpha_{2,c}^*$, $c_1$ and $c_2$ are fitting parameters that depend on $m_c$ and $M$. Lines indicate $\ln(\alpha_{2}^*/\alpha_{2,c}^*) = c_1 \left[ \Delta G_oz(T)/k_{\mathrm{B}}T +c_2 \right]^{2/3}$.}
\end{figure*}

It is a longstanding belief that the fragility of glass formation is somehow related to the extent of dynamic heterogeneity, but any such general relation has remained elusive. We may obtain a direct relation formally through the combination of eq \ref{Eq_String}, which relates $\tau_\alpha$ to $L$, and eq \ref{Eq_NGPTau}, which relates $\alpha_2^*$ to $\tau_\alpha$. We show the result of combining these relations in Figure \ref{Fig_NGP_String}, where we see a scaling of $\alpha_2^*$ with the extent of collective motion $z(T)$ in knotted ring polymer melts. A recent work based on the GET has indicated that the segmental fragility $m$ is directly related to $z(T_\mathrm{g})$ for a wide range of polymer models, \cite{Generalized_2025_21_1664} i.e., $m \approx 7.9 \exp \left [0.6 z(T_\mathrm{g}) \right ]$, so we formally have a relation between the dynamic heterogeneity measure $\alpha_2^*$ and the fragility of glass formation when $\alpha_2^*$ is extrapolated to $T_\mathrm{g}$.

While we do not possess a theoretical framework that allows us to understand exactly in what sense $\alpha_2(t)$ is a measure of dynamic heterogeneity, the above correlations suggest that a deeper understanding of these scaling relationships might offer new and general insights into glass formation. 

\subsection{\label{Sec_Elastic}Elastic Heterogeneity}

Elastic heterogeneity is a phenomenon naturally arising from the expectation that the mobile and immobile regions are associated with long-lived domains exhibiting distinct stiffness characteristics. \cite{Understanding_2022_157_064901} Following the recent works of Zheng et al. \cite{Understanding_2022_157_064901, Competing_2022_55_9990} and Wang et al., \cite{Initiation_2021_155_204504} we adopt eq \ref{Eq_GpDWF} as the defining relationship for local material stiffness. Specifically, to avoid ambiguities inherent in defining a local shear modulus, we utilize $k_{\mathrm{B}} T/ \langle u^2 \rangle$ for individual particles as a measure of ``local stiffness'' of the material. A comprehensive discussion of the quantification of these local elastic heterogeneity fluctuations can be found in the works of Zheng et al. \cite{Understanding_2022_157_064901, Competing_2022_55_9990} 
Notably, nanoscale stiffness fluctuations have been directly observed in nanoprobe measurements on polymer films \cite{Observation_2015_11_1425} and metallic glass materials, \cite{Characterization_2011_106_125504, Relating_2019_7_305} where the qualitative characteristics of these fluctuations closely resemble the results obtained from color maps of local molecular stiffness. In recent works,\cite{Parallel_2023_56_4929, Confinement_2024_160_044503} we also discussed elastic fluctuations in linear and star polymers based on color maps of local molecular stiffness. It is then interesting to examine how knotting affects local molecular stiffness.

\begin{figure*}[htb!]
	\centering
	\includegraphics[angle=0, width=0.95\textwidth]{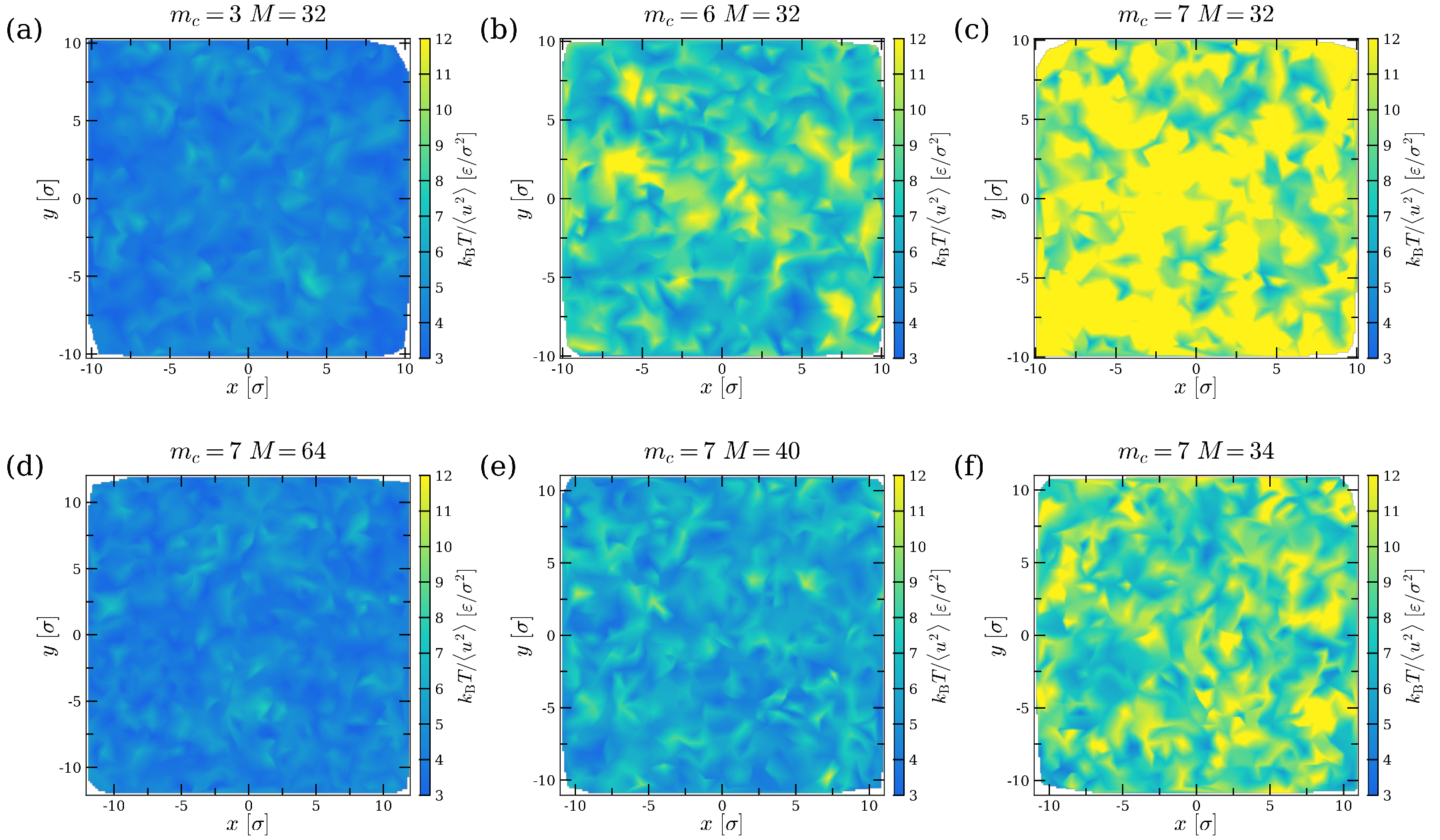}
	\caption{Color maps of local molecular stiffness $k_{\mathrm{B}} T / \langle u^2 \rangle$. Panels (a--f) show the results for various combinations of $m_c$ and $M$ indicated in each panel at $T = 1.0 \varepsilon / k_{\mathrm{B}}$. For purposes of visualization, a slice with a thickness of $2\sigma$ is shown for each system.}
	\label{Fig_Stiffness}
\end{figure*}

Figure \ref{Fig_Stiffness} presents color maps of local stiffness at $T = 1.0 \varepsilon / k_{\mathrm{B}}$ for various combinations of $m_c$ and $M$. Within these maps, regions of relatively high local stiffness appear as yellow, while areas of lower stiffness are shown in blue. To enable direct comparisons in knotted ring polymer melts with varying parameters, a common colorbar scale is used for all systems. These maps reveal that both $m_c$ and $M$ substantially affect the spatial variation in local molecular stiffness. When $m_c$ is relatively small at a fixed $M$, or when $M$ is large at a fixed $m_c$, the color maps display predominantly uniform blue regions, suggesting a nearly homogeneous stiffness distribution at the specified $T$. Under the same thermodynamic conditions, the fluctuations in local stiffness grows progressively as $m_c$ increases at a fixed $M$, and similarly as $M$ increases at a fixed $m_c$. Specifically, increasing either $m_c$ at a fixed $M$ or $M$ at a fixed $m_c$ enhances both the average stiffness and the range of fluctuations in the elastic heterogeneity of the polymer material. This occurs even as the packing efficiency of the molecules increases, resulting in a reduction of free volume as defined by density. This trend is quite reminiscent of the simulation observations on a model glass-forming polymer melt with an anti-plasticizing additive that reduces the fragility and increases the rigidity of the polymer composite material at low $T$. \cite{Antiplasticization_2010_6_292} Our previous study \cite{Glass_2024_57_6875} demonstrated that fragility decreases with increasing $m_c$ when $M$ is small, and similarly, fragility decreases as $M$ decreases for large $m_c$. Therefore, the observed increase in elastic heterogeneity with decreasing fragility is consistent with the established trends in glass-forming liquids. \cite{Antiplasticization_2010_6_292}

\begin{figure*}[htb!]
	\centering
	\includegraphics[angle=0, width=0.8\textwidth]{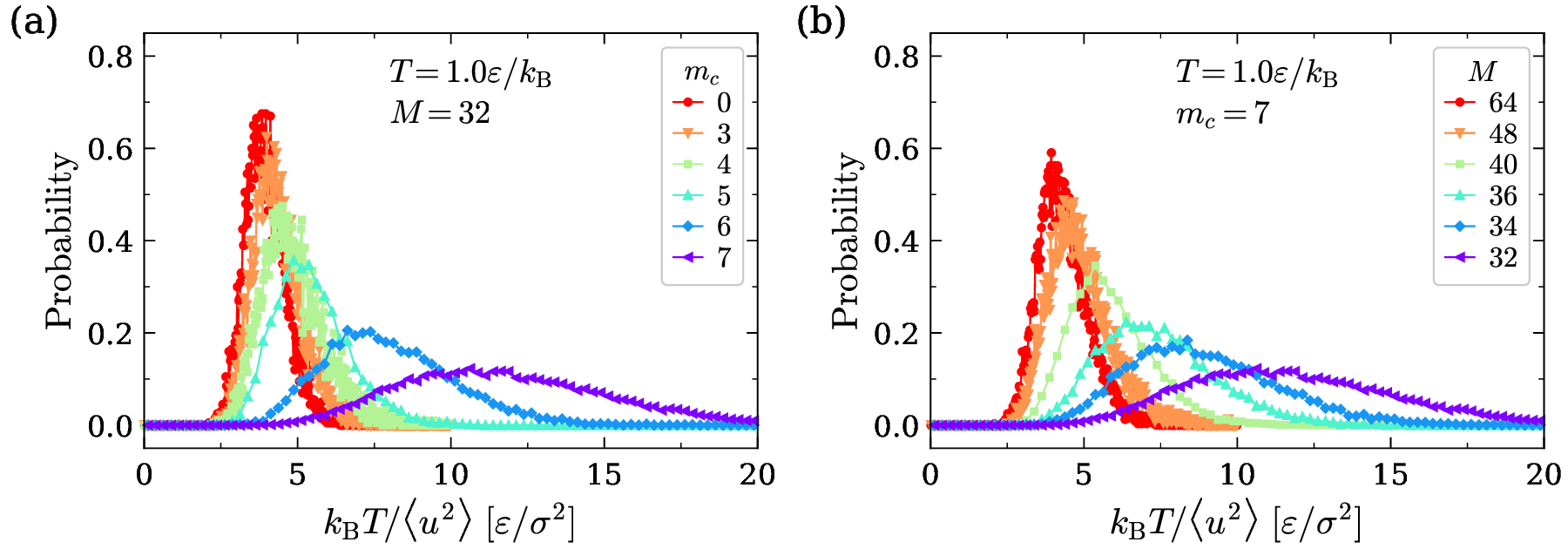}
	\caption{Probability distribution of local molecular stiffness $k_{\mathrm{B}} T / \langle u^2 \rangle$. Panels (a) and (b) show the results for a range of $m_c$ at $M = 32$ and for a range of $M$ at $m_c = 7$ at $T = 1.0 \varepsilon / k_{\mathrm{B}}$, respectively.}
	\label{Fig_PDWF}
\end{figure*}

To further quantify the influence of knotting on local stiffness fluctuations, we analyze the probability distribution of local molecular stiffness, $k_{\mathrm{B}} T / \langle u^2 \rangle$. Figure \ref{Fig_PDWF} shows the results for a range of $m_c$ at fixed $M$ and a range of $M$ at fixed $m_c$ at $T = 1.0 \varepsilon / k_{\mathrm{B}}$. For small $m_c$ or large $M$, it can be seen that a large amount of polymer segments exhibit a low local stiffness and that the shape of the distribution is characterized by a roughly single Gaussian peak, as in previous studies of stiffness fluctuations in simulated polymer materials. \cite{Competing_2022_55_9990, Understanding_2022_157_064901} When $m_c$ is larger than or equal to approximately $6$, the average stiffness of the material, as inferred from $k_{\mathrm{B}} T / \langle u^2 \rangle$, evidently becomes greatly enhanced, since the width of the peak grows progressively with increasing $m_c$ and decreasing $M$. Notably, the probability distribution of local molecular stiffness has an additional peak in star polymers, \cite{Confinement_2024_160_044503} reflecting the segments in the star core regions, but this feature does not appear in knotted ring polymer melts. Overall, this analysis provides another example in which the reduction of fragility, associated with enhanced molecular packing and a reduced fragility of glass formation with higher knot complexity, leads to larger fluctuations in the local stiffness within the material.

Finally, we note that the observed abrupt changes in dynamic and elastic properties around $m_c \approx 5$ at low $M$ can be understood through the lens of topological complexity and geometric constraints. Previous studies on knotted ring, star, and linear polymers have established an approximate correspondence where the properties of knotted rings track those of star polymers with $f \approx m_c + 5$ arms. \cite{Communication_2018_149_161101} In star polymer melts, increasing $f$ beyond approximately $5$ causes a transition toward higher molecular symmetry, and the polymers begin to exhibit particle-like core-shell characteristics as $f$ approaches a value on the order of $10$. Our results indicate that a similar transition occurs in knotted ring melts, where high knot complexity induces a ``particle-like'' character and significantly increases structural rigidity. Crucially, unlike star polymers, knotted rings exhibit a progressive rigidification as their contour length approaches the minimal length required for a given knot topology to exist. In this ``tight knot'' limit, the segments become highly localized, and the structural relaxation time tends to diverge. Thus, the $m_c \approx 5$ threshold marks a crossover from random-coil-like behavior to a regime dominated by topological rigidification and a tendency to form particle-like structures, although the precise value of this threshold probably depends on chain stiffness and intermolecular interaction strength.

\section{Summary}

The enduring relevance of many models of glass formation developed over the years can be attributed to their success in rationalizing observations of real materials. In the present work, we extended our previous efforts to provide a more unified treatment of the dynamics of glass-forming liquids based on extensive simulation data of knotted ring polymer melts, where cooperative motion, rigidity, and glassy dynamics can all be tuned over a wide range. Specifically, we demonstrated that the structural relaxation time can be quantitatively described by the string, localization, and shoving models, which emphasize cooperative particle motion, fluctuations in local particle mobility, and emergent elastic properties, respectively. A primary finding was that despite their superficially different formulations, these theoretical models of glass formation offer essentially equivalent descriptions of relaxation in knotted ring polymer melts. This observation suggested a promising potential for unification in the theories of glass-forming liquids. We also employed the non-Gaussian parameter to characterize dynamic heterogeneity and established the relationship between the peak time of the non-Gaussian parameter, the structural relaxation time, and the glassy plateau shear modulus. We also identified a new relation that links the peak time of the non-Gaussian parameter to the scale of collective motion, thereby connecting this dynamic heterogeneity parameter to the fragility of glass formation. Furthermore, we utilized color maps of local molecular stiffness to visualize the long-lived elastic heterogeneity of knotted ring polymer melts. Overall, our findings reveal a satisfying convergence among distinct theoretical frameworks based on a highly tunable model system. This convergence not only validates the core physical insights embedded in each model but also underscores that the essence of glassy dynamics may be captured by multiple, interrelated physical quantities.



\begin{acknowledgement}
W.-S.X. acknowledges the support from the National Natural Science Foundation of China (Nos. 22573102 and 22222307). This research used resources of the Network and Computing Center at Changchun Institute of Applied Chemistry, Chinese Academy of Sciences. The authors thank Dr. Fernando Vargas-Lara at ExxonMobil Technology and Engineering Company for sharing the initial knot configurations and helpful discussions about knotted rings. W.-S.X. and Y.-T.D. are grateful to Prof. Liang Dai at City University of Hong Kong for useful discussions about knotted rings and for sharing the code for identifying knot types.
\end{acknowledgement}

\begin{suppinfo}
Definition of basic dynamic properties, additional results about correlations between characteristic properties of glass formation, and influence of the caging onset time on model fits.
\end{suppinfo}

\bibliography{refs}

\end{document}